\documentclass[article]{article}
\usepackage{color}
\usepackage{graphicx}
\usepackage{amsmath}
\numberwithin{equation}{section}
\usepackage{amssymb}
\usepackage{mathrsfs}
\usepackage[numbers,sort&compress]{natbib}
\usepackage{amsthm}
\usepackage{tikz}
\begin{document}
	
	\title{Long time  asymptotics behavior of the focusing  nonlinear Kundu-Eckhaus equation}
	\author{ Ruihong Ma and Engui Fan  }
	\footnotetext[1]{ \  School of Mathematical Sciences, Furan University, Shanghai 200433, P.R. China.}

	\date{ }
	\maketitle
	\begin{abstract}
		\baselineskip=17pt
		We study the Cauchy problem for the focusing nonlinear  Kundu-Eckhaus  equation and  construct   long time asymptotic expansion of its solution  in  fixed space-time cone with  $C(x_1,x_2,v_1,v_2)=\{(x,t)\in\Re^2:x=x_0+vt$ $x_0\in[x_1,x_2],v\in[v_1,v_2] \}$.  By  using the inverse scattering transform, Riemann-Hilbert approach  and  $\overline\partial$ steepest descent method we   obtain the lone time asymptotic behavior of the solution, 
at the same time  we obtain the solitons in the cone compare with the all N-soliton  the  residual error up to order $\mathcal{O}(t^{-3/4})$.
		\\
		\\Keywors:    Focusing   Kundu-Eckhaus equation;  Lax pair;  Riemann-Hilbert problem; $\overline\partial$ steepest descent method;  soliton solution; long time asymptotics.
	\end{abstract}

	\baselineskip=17pt
	
	\newpage
	
	\section{\noindent\textbf\expandafter{ Introduction}
		\\\hspace*{\parindent}\vspace{2mm}}
	$\qquad$ We study the long time asymptotic behavior of the focusing nonlinear  Kundu-Eckhaus(fKE) equation on $\Re\times\Re_{+}$:
	\begin{align}
&	iq_t+q_{xx}+2|q|^2q+4\beta^2|q|^4q-4i\beta(|q|)^2_xq=0,\label{eq:1.1}\\
& q(x,0)=q_0(x).
	\end{align}

In the defocusing case   with  the sign of cubic term reversed
\begin{equation}\label{eq:1.1}
	iq_t+q_{xx}-2|q|^2q+4\beta^2|q|^4q-4i\beta(|q|)^2_xq=0,
	\end{equation}
   and  initial value $q_0(x)$   in Schwarz space,  it has known  that \cite{[1]} as $t\to\infty$
	\begin{equation}\label{eq:1.2}
	q(x,t)=\sqrt{\frac{\nu}{2t}}e^{i(4tk_{0}^{2}-\nu(k_{0})\log 8t)}e^{i\phi(k_{0})}e^{-\frac{2}{\pi}i\beta\int_{k_0}^{\infty}\log(1-|r^{\prime}(k^{\prime})|^2)
dk^{\prime}}+\mathcal{O}(t^{-1}\log t),
	\end{equation}
	where
	\begin{equation}\label{eq:1.3}
	k_{0}=-\frac{x}{4t},
	\end{equation}
	\begin{equation}\label{eq:1.4}
	\nu(k_{0})=-\frac{1}{2\pi}\log(1-|r(k_{0})|^2,
	\end{equation}
		\begin{equation}\label{eq:1.5}
	\phi(k_{0})=\frac{1}{\pi}\int_{-\infty}^{k_{0}}\log(k_{0}-k^{\prime})d \log(1-|r(k^{\prime}|^2)+\frac{\pi}{4}-\arg r(k_{0})+\arg\Gamma(i\nu),
	\end{equation}
	where $\Gamma$  is Gamma function,  $r(z)$ is the reflection coefficient.  
	Recently, We   applied Riemann-Hilbert approach to
$N$-soliton solutions for the fKE equation (\ref{eq:1.1}) with  nonzero boundary conditions \cite{fan2019}. 
 Wang et al obtained long-time asymptotics of the fKE equation (\ref{eq:1.1}) with  nonzero boundary conditions \cite{wang2018}.
In this paper,  we consider a much weaker boundary condition  that suppose  $q_0(x)$ in the weighted Sololev space
	\begin{equation}\label{eq:1.6}
	H^{1,1}=\{f \in L^{2}(\Re) : xf , f^{\prime}\in L^{2}(\Re)\}.
	\end{equation}
There exist a nonzero complex number $c_k$ called a norming constant associated with any point in the simple discrete spectrum $z_k\in\mathbb{C}^+$. Define reflection coefficient $r:\Re\to\mathbb{C}$
(where in the ZS-AKNS operator we know that the real axis is the continuous spectrum),
and reflection coefficient $r$ may take any value in $\mathbb{C}$ in the focusing case \cite{[2]}.  If $r$ has singularities along the real line, call it spectral singularities; if there exist spectral singularities, it is possible that infinite discrete spectrum  accumulate at a spectral singularity\cite{[9]}.  In this paper, we only consider that no spectral singularities exist so the discrete spectrum is finite.
	If the spectrum consist of a single point, $\sigma_{d}=\{(\xi+i\eta)\}$ the corresponding solution of (1.1), is the one soliton
	\begin{equation}\label{eq:1.7}
	q_{sol}(x,t)=q_{sol}(x,t;\{\xi+i\eta\})=2ia_{1}e^{\Omega_{1}-\Omega_{1}^{*}}(P^{-1})e^{8i\beta\int |ae^{\Omega_{1}-\Omega_{1}^{*}}(P^{-1})^{2}|dx},
	\end{equation}
	\begin{equation}\label{eq:1.8}
	v_{1}=e^{-iz_{1}\sigma_{3} x-2iz_{1}^{2}\sigma_{3}tv}v_{10},
	\end{equation}
	\begin{equation}\label{eq:1.9}
	\widetilde v_{1}=\widetilde v_{10}e^{i\widetilde z_{1}\sigma_{3} x+2i\widetilde z_{1}^{2}\sigma_{3}tv},
	\end{equation}
	\begin{equation}\label{eq:1.10}
	P=\frac{\widetilde v_{1}v_{1}}{z^{*}_{1}-z_{1}},
	\end{equation}
where $\Omega_{1}=-iz_{1}x-2iz_{1}^{2}t $ and $a_{1} $ and $z_{1}$ are complex constants and $v_{10} $ is initial speed.
Let  $q_{sol}(x,t;\sigma _{d}) $ stand for  $N$-soliton solution with  scattering data $\{r\equiv0,\sigma_{d}=\{(z_{k},c_{k})_{k=1}^{N}\}$.
Generally, the solution breaks apart into $N$ independent one-soliton, each traveling at initial speed $v_{k}$ \cite{[4]}.
\\
	\emph{1.1. Main results and remark}\\
\hspace*{\parindent}In order to describes the asymptotic behavior of the solution (\ref{eq:1.1})  as $t\to\infty $, for generic intimal data $q_0\in H^{1,1}(\Re)$. Define the discrete scattering data $\{r,\{(z_{k},c_{k})\}_{k=1}^{N}\}$. Let $\mathcal{Z}=\{z_{k}\}_{k=1}^{N}\subset  \mathbb{C}^{+}$.Define
\begin{equation}\label{eq:1.11}
	\kappa(s)=-\frac{1}{2\pi}log(1+|r(s)|^{2}),
	\end{equation}
	and for any number $\xi$,  let
	\begin{eqnarray}\label{eq:1.12}
	\begin{split}
	&\bigtriangleup_{\xi}^{-}=\{k\in\{1,2,\cdots,N\}:Rez_{k}<\xi\},\\
	&\bigtriangleup_{\xi}^{+}=\{k\in\{1,2,\cdots,N\}:Rez_{k}>\xi\},
	\end{split}
	\end{eqnarray}
	Given any real interval $\mathcal{I}=[a,b]$, let
	\begin{eqnarray}\label{eq:1.13}
	\begin{split}
	&\mathcal{I}=|\mathcal{Z}(\mathcal{I})|,\\
	&\mathcal{Z}(\mathcal{I})=\{z_{k}\in\mathcal{Z}:Re z_{k}\in\mathcal{I}\},\\
	&\mathcal{Z}^{-}(\mathcal{I})=\{z_{k}\in\mathcal{Z}:Re z_{k}<a\},\\
	&\mathcal{Z}^{+}(\mathcal{I})=\{z_{k}\in\mathcal{Z}:Re z_{k}>b\},
	\end{split}
	\end{eqnarray}
	For $\xi\in\mathcal{I}$,  let
	\begin{eqnarray}\label{eq:1.14}
	\begin{split}
	&\Delta_{\xi}^{-}=\{k\in\{1,2,\cdots ,N\}: a\leq{Rez_{k}}<\xi\},\\
	&\Delta_{\xi}^{+}=\{k\in\{1,2,\cdots ,N\}:\xi<Re z_{k}\leq {b}\},\\
	&\sigma_{d}^{\pm}=\{(z_{k},c_{k}^{\pm}(\mathcal{I}):z_{k}\in\mathcal{I})\},\\
	&c_{k}^{\pm}=c_{k}\prod_{z_{j}\in\mathcal{Z}^{\pm}(\mathcal{I})}\Bigg(\frac{z_{k}-z_{j}}{z_{k}-z_{j}^{*}}\Bigg)^{2}exp \Bigg(\pm2i\int_{\xi}^{\mp\infty}\frac{\kappa(s)}{s-z_{k}}ds\Bigg),
	\end{split}
	\end{eqnarray}
Finally, given pairs of velocities $v_{1}\leq v_{2}$ and points $x_{1}\leq x_{2} $ define the cone
	\begin{equation}\label{eq:1.15}
	C(x_{1},x_{2},v_{1},v_{2}):=\left\{(x,t)\in\Re^{2}:x=x_{0}+vt\quad
	with  \quad     x_{0}\in[x_{1},x_{2}],v\in[v_{1},v_{2}]\right\}),
	\end{equation}
	\begin{tikzpicture}[node distance=2cm]
	\draw[->](-3,0)--(5,0) node[right] {$Rez$};
	\draw[->](0,-1)--(0,5);
	\draw[->](2,-1)--(2,5);
	\draw(0,0)--(0,0.1) node[below]{$-v_2/2$};
	\draw(2,0)--(2,0.1) node[below]{$-v_1/2$};
	\draw(-2,1)--(-2,1.1) node[below]{$z_7$};
	\draw(-1,3)--(-1,3.1) node[below]{$z_5$};
	\draw(-0.5,3.5)--(-0.5,3.6) node[below]{$z_4$};
	\draw(0.5,2)--(0.5,2.1) node[below]{$z_6$};
	\draw(1,1)--(1,1.1) node[below]{$z_8$};
	\draw(1.5,3)--(1.5,3.1) node[below]{$z_3$};
	\draw(3,1)--(3,1.1) node[below]{$z_9$};
	\draw(2.5,3)--(2.5,3.1) node[below]{$z_2$};
	\draw(4,3.7)--(4,3.8) node[below]{$z_1$};
	\end{tikzpicture}
	\begin{tikzpicture}[node distance=2cm]
	\draw[->](-3,0)--(3,0) node[right] {$x$};
	\draw[->](0,-3)--(0,3) node[above]{$t$};
	\draw(1,0)--(1,0.1) node[below]{$x_1$};
	\draw(2,0)--(2,0.1) node[below]{$x_2$};
	\draw(1,0)--(0.5,3)node [left]{$x-v_1t=x_1$};
	\draw(2,0)--(3,3)node [above]{$x-v_2t=x_2$};
	\draw(1,0)--(-1,-3)node [left]{$x-v_2t=x_1$};
	\draw(2,0)--(3,-3)node [above]{$x-v_1t=x_2$};
	\end{tikzpicture}
	\\
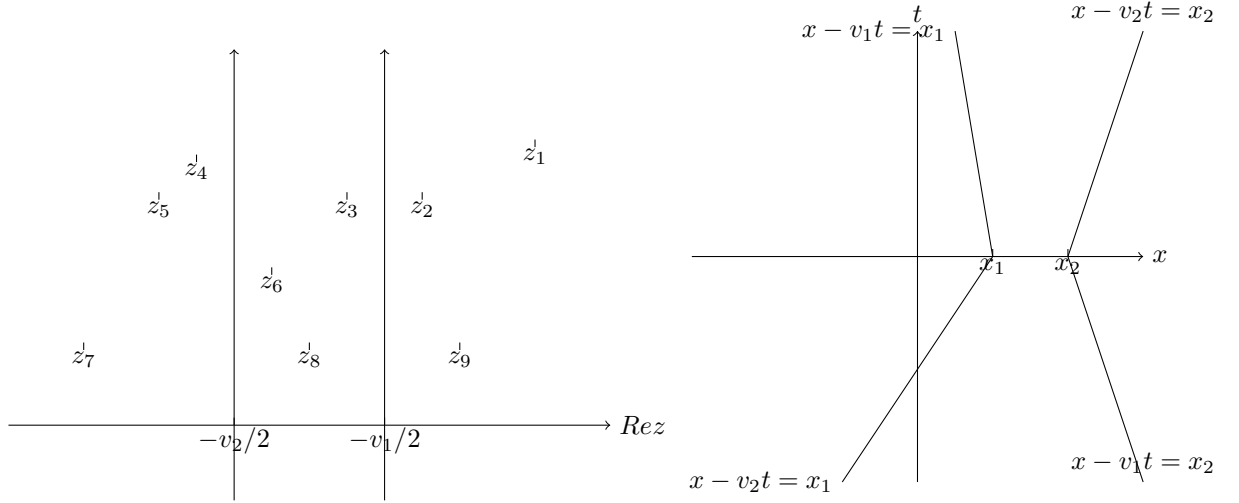
	\emph{ Fig.1.1:Using soliton contained in the cone C associated with its reflectionless scattering data  to describe the asymptotic behavior of $q(x,t)$ as $|t|\to\infty$.In the example here,we use 3-soliton
		$\mathcal{Z}(\mathcal{I})=\{z_3,z_6,z_8\}$ those inside the cone  C asymptotically described the  discrete spectrum instead of 9-soliton}
\\
\noindent\textbf{Assumption 1.1.} The initial data $q_0(x)$ for the Cauchy problem of fKE satisfied the  scattering dates as followed:
	\\
1)	Every $z_k\in\mathbb{C}^+$ satisfied $a(z_k)=0$ is simple that is the discrete spectrum is simple.
\\
2)There exist a constant $c>0$ such that $|a(z)|\geq$0 that is no spectral singularities exist.
\\
According to above assumption which guarantees that the discrete spectrum is finite.
\\
\noindent\textbf{Theorem 1.1.}  Let  $q(x,t)$ be the solution of (\ref{eq:1.1}) satisfied with initial data $q_{0}(x)\in H^{1,1}(\Re)$ ,suppose it satisfies Assumption 1.1,and generate the scattering data  $\{r,\{z_{k},c_{k}\}_{k=1}^{N}\}$ ,Fix $x_{1},x_{2},v_{1},v_{2}\in\Re$ with $x_{1}<x_{2}$ and $v_{1}<v_{2}$ .Let $\mathcal{I}=[-\frac{v_{2}}{2},-\frac{v_{1}}{2}]$,and let $\xi=-\frac{x}{4t}$.Then, when  $t\to\pm\infty$ with $(x,t)\in C(x_{1},x_{2},v_{1},v_{2})$,with $C$ as define in $(1.15)$,we have
	\begin{equation}\label{eq:1.16}
	q(x,t)=(q_{sol}(x,t;\sigma^{\pm}(I))+t^{-\frac{1}{2}}f^{\pm}(x,t)+O(t^{-\frac{3}{4}}))e^{-8i\beta\int_{(-\infty,t)}^{(x,t)}|m|^{2}dx}+\mathcal{O}({t^{-\frac{3}{4}}}),
	\end{equation}
	\begin{equation}\label{eq:1.17}
	f^{\pm}(x,t)=m_{11}(\xi;x,t)^{2}\alpha(\xi,\pm)e^{ix^{2}/(2t)\mp i\kappa(\xi)log|4t|}+m_{12}(\xi;x,t)^{2}\alpha(\xi,\pm)^{*}e^{-ix^{2}/(2t)\pm i\kappa(\xi)log|4t|},
	\end{equation}
	with
	\begin{equation}\label{eq:1.18}
	|\alpha(\xi,\pm)|^{2}=|\kappa(\xi)|,
	\end{equation}
	\begin{equation}\label{eq:1.19}
	arg\alpha(\xi,\pm)=\pm\frac{\pi}{4}\pm arg\Gamma(i\kappa(\xi))-arg r(\xi)-4\operatorname*{\sum}\limits_{k\in\Delta_{\xi}^{\mp}}arg(\xi-z_{k})\mp2\int_{\mp\infty}^{\xi}log|\xi-s|d_{s}\kappa(s),
	\end{equation}
	\begin{equation}\label{eq:1.20}
	|m(x,t)|^2=\Bigg|\frac{1}{2i}(q_{sol}(x,t;\sigma^{\pm}(I))+t^{-\frac{1}{2}}f^{\pm}(x,t)+\mathcal{O}(t^{-\frac{3}{4}}))\Bigg|^2,
	\end{equation}
	the coefficients $m_{11}(\xi;x,t)$and $m_{12}(\xi;x,t)$ are the entries in the first row of the solution of RHP A.2 with discrete scattering data $\sigma_{d}^{\pm}(I)$and $\Delta=\Delta_{\xi}^{\mp}(I)$ evaluated at $a=\xi$.
	\\
	
\emph{1.2. Organization of the rest of the paper}

In the   Section 2,  we first construct  the  RHP 2.1  associated with initial-value problem (\ref{eq:1.1}),  and then   we  work out   the steepest descent analysis of RHP 2.1 for
 $t\to\infty$ from  Section 3 to Section 7.   In Section 3 we introduce the matrix $T(z)$ to separate the jump matrix defined in RHP 2.1 at $\xi=-\frac{x}{4t}$,  In Section 4,  introduces the $\overline\partial$ analysis to define extensions of the jump matrix for the non-linear steepest descent method.  In section 5 we construct a global model solution which captures the leading order asymptotic behavior of the solution.Removing this component of the solution results in a small norm $\overline\partial$- problem which is analyzed in Section 6.The proof of Theorem 1.1 is given in Section 7.
	\section{\noindent\textbf{\expandafter{ Results of scattering theory for focusing KE }\\\hspace*{\parindent}\vspace{2mm}}}
	
	Our calculations are based on the Lax pair of the focusing KE equation (\ref{eq:1.1})
	\begin{equation}\label{eq:2.1}
	\psi_{x}+iz\sigma_{3}\psi=Q_{1}\psi,
	\end{equation}
	\begin{equation}\label{eq:2.2}
	\psi_{t}+2iz^2\sigma_{3}\psi=Q_{2}\psi,
	\end{equation}
	where $z$  is a spectral parameter and
	\begin{eqnarray}\label{eq:2.3}
	\begin{split}
	&Q=\begin{pmatrix}
	0&q\\
	-\overline{q}&0
	\end{pmatrix},
	\sigma_{3}=\begin{pmatrix}
	1&0\\
	0&-1
	\end{pmatrix},\\
	&Q_{1}=Q-i\beta Q^{2}\sigma_{3},\\
	&Q_{2}=4i\beta^{2}Q^{4}\sigma_{3}-2\beta Q^{3}-iQ^{2}\sigma_{3}+2zQ-iQ_{x}\sigma{3}+\beta(Q_{x}Q-QQ_{x}),\\
	\end{split}
	\end{eqnarray}
	Specially, z$ \in\Re$ ,then the eigenfunction $\psi(x,t,z)$ in the Lax pair ,letting:
	$$\Psi=\psi e^{i(zx+2z^{2}t)\sigma_3},$$
	obtain the equivalent Lax pair
	\begin{eqnarray}\label{eq:2.4}
	\begin{split}
	&\Psi_{x}+iz[\sigma,\Psi]=Q_{1}\Psi,\\
	&\Psi_{t}+2iz^{2}[\sigma_{3},\psi]=Q_{2}\Psi,
	\end{split}
	\end{eqnarray}
	its full derivative form is
	\begin{equation}\label{eq:2.5}
	d(e^{i(zx+2z^{2}t})\hat{\sigma_{3}}\Psi(x,,t,k))=e^{i(zx+2z^{2}t)\hat{\sigma_{3}}}U\Psi,
	\end{equation}
	\begin{equation}\label{eq:2.6}
	U=Q_{1}dx+Q_{2}dt,
	\end{equation}
	consider a solution of (\ref{eq:2.5})of the form
	\begin{equation}\label{eq:2}
	\Psi=D+\frac{\Psi_{1}}{z}+\frac{\Psi_{2}}{z^{2}}+O(\frac{1}{z^{3}}),z\to\infty,
	\end{equation}
where $D,\Psi_{1},\Psi_{2}$ are independent of $z$,substituting  above expansion into two equations of (\ref{eq:2.4}),and comparing the same order of $z^\prime$ frequency,we find the following equations
	\begin{equation}\label{eq:2.7}
	D_{x}=-i\beta Q^{2}\sigma_3D,
	\end{equation}
	\begin{equation}\label{eq:2.8}
	D_{t}=(\beta(u\overline{u}_{x}-u_{x}\overline{u}+4i\beta^{2}|u|^{4})\sigma_3D,
	\end{equation}
	we get  (\ref{eq:1.1}) has the conservation law
	$$(i\beta|u|^{2})_{t}=(\beta(u\overline{u}_{x}-u_{x}\overline{u})+4i\beta^{2}|u|^{4})_{x},$$
	the two (\ref{eq:2.7}) and (\ref{eq:2.8}) for $D$ are consistent and are both satisfied if we define
	\begin{equation}\label{eq:2.9}
	D(x,t)=e^{i\int_{(x,t)}^{(+\infty,t)}\Delta\sigma_{3}},
	\end{equation}
	where $\Delta$ is
	\begin{equation}\label{eq:2.10}
	\Delta(x,t)=\beta|u|^{2}dx+(-i\beta(u\overline{u}_{x}-u_{x}\overline{u})+4\beta^{2}|u|^{4})dt,
	\end{equation}
According to asymptotic analysis we introduce a new function $\mu$ by
	\begin{equation}\label{eq:2.11}
	\Psi(x,t,z)=e^{-i\int_{(-\infty,t)}^{(x,t)}\Delta\hat\sigma}\mu(x,t,z)D(x,t),
	\end{equation}
	Thus ,we have
	\begin{equation}\label{eq:2.12}
	\mu=I+O(\frac{1}{z}),z\to\infty,
	\end{equation}
	and (\ref{eq:2.5}) becomes
	\begin{equation}\label{eq:2.13}
	d(e^{i(zx+2z^{2}t)\hat{\sigma_{3}}}\mu(x,t,k))=W(x,t,z)=e^{i(zx+2z^{2}t)\hat{\sigma_{3}}}V(x,t,z)\mu,
	\end{equation}
	with
	\begin{equation}\label{eq:2.14}
	V=V_{1}dx+V_{2}dt,
	\end{equation}
	with
	\begin{equation}\label{eq:2.15}
	V_{1}=\begin{pmatrix}
	2i\beta|u|^2&ue^{2i\int_{(-\infty,t)}^{(x,t)}}\Delta\\
	-\overline{u}e^{-2i\int_{(-\infty,t)}^{(x,t)}}&-2i\beta|u|^2
	\end{pmatrix},
	\end{equation}
	\begin{equation}\label{eq:2.16}
	V_{2}=\begin{pmatrix}
	i|u|^2&(2\beta u|u|^2+2zu+iu_x)e^{2i\beta\int_{(-\infty,t)}^{(x,t)}}\Delta\\
	(-2\beta\overline{u}|u|^2-2z\overline{u}-i\overline{u}_x)e^{-2i\beta\int_{(-\infty,t)}^{(x,t)}}\Delta&-i|u|^2
	\end{pmatrix},
	\end{equation}
	The Lax pair (\ref{eq:2.3})can be change into
	\begin{eqnarray}\label{eq:2.17}
	\begin{split}
	&\mu_{x}+iz[\sigma_3,\mu]=V_{1}\mu,\\
	&\mu_{t}+2iz^{2}[\sigma_{3},\mu]=V_{2}\mu,
	\end{split}
	\end{eqnarray}
	We assume that $\mu(x,t)$ is sufficiently smooth,we define two solutions of (\ref{eq:2.13}) by
	\begin{equation}\label{eq:2.18}
	\mu_{j}(x,t,z)=I+\int_{(x_{j},t_{j})}^{(x,t)}e^{-i(zx+2z^{2}t)\hat{\sigma}_{3}}W(y,\tau,z)dy,\qquad j=1,2,
	\end{equation}
where $(x_{1},t_{1})=(-\infty,t),(x_{2},t_{2})=(+\infty,t)$
it follows that $det \Psi_{(1,2)}=det\mu_{(1,2)}\equiv1$.
and it satisfied
\begin{equation}\label{eq:2.19}
\mu_{x}+iz[e^{-i\int_{(x,t)}^{(+\infty,t)}\Delta\sigma_{3}}
\sigma_{3},\mu]=Q\mu,
\end{equation}
that is
\begin{equation}\label{eq:2.20}
\mu_{x}+iz[D^{-1}\sigma_{3},\mu]=Q\mu,
\end{equation}
its Volterra type integrals
\begin{equation}\label{eq:2.21}
\mu_{(1,2)}(x,z)=I+\int_{\pm\infty}^xe^{izD^{-1}(x-y)\sigma_{3}}Q\mu_{(1,2)}(y,z)e^{-izD^{-1}(x-y)\sigma_{3}}dy,
\end{equation}
Also,if $\mu(x,z)$ is any solution of (\ref{eq:2.17}) then $\tilde{\mu}(x,z)=\sigma_2\overline{\mu(x,\overline{z})}\sigma_2$(complex conjugate but no transpose) also solve (\ref{eq:2.17}).For $z\in\Re$ ,$\sigma_2\overline{\mu_{(1,2)}(x,\overline{z})}\sigma_2$ also satisfies (\ref{eq:2.17}) and it follows that
	\begin{equation}\label{eq:2.22}
	\Psi_{(1,2)}(x,z)=\sigma_2\overline{\Psi_{(1,2)}(x,\overline{z})}\sigma_2,\quad z\in\Re,
	\end{equation}
Since the eigenfunctions  $\mu_{1}(x,t,z)$ and $\mu_{2}(x,t,z)$ satisfy both equations of the Lax pair (\ref{eq:2.17}) ,so there exists a continuous scattering matrix function $S(z)$ satisfying
	\begin{equation}\label{eq:2.23}
	\mu_{1}(x,t,z)=\mu_{2}(x,t,z)e^{-i(zx+2z^{2}t)\hat{\sigma_{3}}}S(z),\quad z\in\Re,
	\end{equation}
	\begin{eqnarray}\label{eq:2.24}
	\begin{split}
	&S(z)=\begin{pmatrix}
	a(z)&-\overline{b(\overline{z})}\\
	b(z)&\overline{a(\overline{z})}
	\end{pmatrix},
	&detS(z)=|a(z)|^{2}+|b(z)|^{2}=1,\\
	\end{split}
	\end{eqnarray}
Define
	\begin{eqnarray}\label{eq:2.25}
	\begin{split}
	&\mu_{1}=(\mu_{1}^{(1)},\mu_{1}^{(2)})=
	\begin{pmatrix}
	\mu_{1}^{(11)}&\mu_{1}^{(12)}\\
	\mu_{1}^{(21)}&\mu_{1}^{(22)}
	\end{pmatrix}
	&&\mu_{2}=(\mu_{2}^{(1)},\mu_{2}^{(2)})=
	\begin{pmatrix}
	\mu_{2}^{(11)}&\mu_{2}^{(12)}\\
	\mu_{2}^{(21)}&\mu_{2}^{(22)}
	\end{pmatrix}
	\end{split},
	\end{eqnarray}
where  $\mu_{j}^{1}(x,t,z) $ and $\mu_{j}^{2}(x,t,z)$
are the first and second columns of $\mu_{j}(x,t;z)$, $j=1,2$

\noindent\textbf{Remark 2.1 }
\\
$\bullet $$ \mu_{1}^{(1)},\mu_{2}^{(2)}$ and $a(z)$ extend analytically to $z\in\mathbb{C}^+$ with continuous boundary values on $\Re$,and $\mu_{1}^{(1)}\to e_1,\mu_{2}^{(2)}\to e_2$ and $a(z)\to1$ when $z\to\infty$,similarly consequence hold for $z\in\mathbb{C}^-$,however,$b(z)$ is defined only for $z\in\Re$.
\\
$\bullet$ the solutions $\Psi_1^{(1)}(x,z_k) $and $\Psi_2^{(2)}(x,z_k)$ are linearly dependent when $a(z_k)=0$ for  $z_k\in\mathbb{C}$.so there exist a norming constants  $c_k$ satisfied
	\begin{equation}\label{eq:2.26}
	\Psi_1^{(1)}(x,z_k)=c_k\Psi_2^{(2)}(x,z_k),
	\end{equation}
	The symmetry (\ref{eq:2.19})) implies that
	\begin{equation}\label{eq:2.27}
	\Psi_2^{(1)}(x,z_k^*)=c_k^*\Psi_1^{(2)}(x,z_k^*),
	\end{equation}
$\bullet$ The reflection coefficient  $ r $ and transmission coefficient  $ \tau$ are defined by
	\begin{equation}\label{eq:2.28}
	r(z)=\frac{b(z)}{a(z)}\qquad \tau=\frac{1}{a(z)},
	\end{equation}
	and it follows from (\ref{eq:2.21}) that $1+|r(z)|^2=|\tau(z)|^2$ for each $z\in\Re$
	\\
We construct  the function
	\begin{equation}\label{eq:2.29}
	M(z)=M(z;x,t):
	\begin{cases}
	\Bigg[\frac{\mu_{1}^{(1)}(x,t;z)}{a(z)}  \qquad \mu_{2}^{(2)}(x,t;z)\Bigg] ,when z\in\{z\in \mathbb{C}|Im z>0\},\\
	\\
	\Bigg[\mu_{2}^{(1)}(x,t;z) \qquad \frac{\mu_{1}^{(2)}(x,t;z)}{\overline{ a(\overline z)}}\Bigg],when z\in\{z\in \mathbb{C}|Im z<0\},
	\end{cases}
	\end{equation}
The matrix $M$ defined above is the solution of the following Riemann-Hilbert problem.
	\\
	\noindent\textbf{Riemann-Hilbert Problem 2,1 }Fine an analytic function $M:\mathbb{C}\setminus(\Re \cup \mathcal{Z} \cup \mathcal{Z}^{*})\to SL_{2}(\mathbb{C})$ with the following properties
	\begin{enumerate}
		\item $M(z)=I+O(z^{-1})\quad as\quad z \to\infty $
		\item The  continuous boundary values $M_{\pm}(z)$  satisfy the jump relation $M_{+}(x,t,z)=M_{-}(x,t,z)J(x,t,z)$, $z\in\Re$ where
		\begin{equation}\label{eq:2.31}
		J(z)=\begin{pmatrix}
		1+|r(z)|^{2}&r^{*}(z)e^{-2it\theta(z)} \\
		r(z)e^{2it\theta(z)}&1
		\end{pmatrix},
		\end{equation}
		\begin{equation}\label{eq:2.32}
		\theta(z):=2z^{2}+z\frac{x}{t}=2(z-\xi)^2-2\xi^2\qquad\xi=-\frac{x}{4t},
		\end{equation}
		\item $M(z)$ has simple poles at each $z_{k}\in \Re $ and  $ z_{k}^{*}\in\Re^{*}$
		at which
		\begin{eqnarray}\label{eq:2.33}
		\begin{split}
		&\operatorname*{Res}\limits_{z_{k}}M=\operatorname*{lim}\limits_{z\to z_{k}}M\begin{pmatrix}
		0&0\\
		c_{k}e^{2it\theta}&0
		\end{pmatrix},
		\\
		&\operatorname*{Res}\limits_{z_{k}^{*}}M=\operatorname*{lim}\limits_{z\to z_{k}^{*}}M\begin{pmatrix}
		0&-c_{k}^{*}e^{-2it\theta}\\
		0&0
		\end{pmatrix},
		\end{split}
		\end{eqnarray}
	\end{enumerate}	
Consider the (\ref{eq:2}),we get $\Psi_1=\frac{i}{2}QD\sigma_3$,
$i[\sigma_{3},\Psi_{1}]=QD$.
The existence of solutions of  RHP 2.1 for all $(x,t)\in \Re^{2}$ follows by means of expanding this solution as
$z\to\infty,M=I+\frac{M^{(1)}(x,t)}{z}+o(z^{-1})$,one finds that
	\begin{equation}\label{eq:2.34}
	q(x,t)=\operatorname*{lim}\limits_{z\to\infty}(2izM(x,t,z))_{12}e^{-2i\int_{(-\infty,t)}^{(x,t)}\Delta}=2im(x,t)e^{-2i\int_{(-\infty,t)}^{(x,t)}}\Delta,
	\end{equation}
where
\begin{equation}\label{eq:2.35}
m(x,t)=\operatorname*{lim}\limits_{z\to\infty}(zM(x,t,z))_{12},
\end{equation}
	And
	\begin{equation}\label{eq:2.36}
	\mu=I+\frac{\mu^{(1)}}{z}+\frac{\mu^{(2)}}{z^2}+\mathcal{O}(\frac{1}{z^3}),\quad z\to\infty,
	\end{equation}
	is the corresponding solution of (\ref{eq:2.13}) related to $\Psi$ via (\ref{eq:2.11}),From its complex conjugate,we obtain
	$$|q|^2=4|m|^2,$$
	$$u\overline{u}_x-u_x\overline{u}=4(m\overline{m}_x-\overline{m}m_x)+64i\beta|m|^4,$$
	Thus,we are able to express the one-form $\Delta$ defined in (\ref{eq:2.34}) in terms of $m$ as
	\begin{equation}\label{eq:2.37}
	\Delta=4\beta|m|^2dx+[4i\beta(\overline{m}m_x-m\overline{m}_x)+128\beta^2|m|^4]dt,
	\end{equation}
\section{\noindent\textbf{\expandafter{ Conjugation }\\\hspace*{\parindent}\vspace{2mm}}}
In this section,we introduce the function $T(z)$ to renormalized the Riemann-Hilbert problem  with $\xi$ fixed
	\begin{eqnarray}\label{eq:3.1}
	\begin{split}
	&T(z)=T(z,\xi)=\operatorname*{\prod}\limits_{k\in\bigtriangleup_{\xi}^{-}}\Bigg(\frac{z-z_{k}^{*}}{z-z_{k}}\Bigg)exp\Bigg(i\int_{-\infty}^{\xi}\frac{\kappa(s)}{s-z}ds\Bigg),
	\\
	&\kappa(s)=-\frac{1}{2\pi}log(1+|r(s)|^{2}),\\
	\end{split}
	\end{eqnarray}
we can also get the standard result of the transmission coefficient
	\begin{equation}\label{eq:3.2}
	\frac{1}{a(z)}=\prod_{k=1}^{N}\Bigg(\frac{z-z_k^*}{z-z_k}\Bigg)exp\Bigg(i\int_{-\infty}^{\infty}\frac{\kappa{(s)}}{s-z}ds\Bigg),
	\end{equation}
	and we can find  $T(Z;\xi)\to1/a(z)$ when $\xi\to\infty$
\\\noindent\textbf{\expandafter {Proposition3.1}} The function $T(z)$ defined by  (\ref{eq:3.1}) has the following prosperities:
\\
(a)T  is nonzero and meromorhpic in  $\mathbb{C}\setminus(-\infty,\xi]$,For each k in $\Delta_{\xi}^{-}$, $T(z)$ has a simple pole at $z_{k}$ and a simple zero at $\overline{z_{k}}$.
\\
	(b) For $z\in\mathbb{C}\setminus (-\infty,\xi]$,$\overline{T(\overline{z})}=1/T(z)$
\\
	(c)For $z\in(-\infty,\xi]$,the boundary values $T_{\pm}$sarisfy
	\begin{equation}\label{eq:3.3}
	T_{+}(z)/T_{-}(z)=1+|r(z)|^{2},\quad z\in(-\infty,\xi),
	\end{equation}
\\
	(d) As $|z|\to\infty$ with$ |arg(z)|\leq c\leq\pi,$
	\begin{equation}\label{eq:3.4}
	T(z)=1+\frac{i}{z}\Bigg[2\operatorname*{\sum}\limits_{k\in\Delta_{\xi}^{-}}Im z_{k}-\int_{-\infty}^{\xi}\kappa(s)ds\Bigg]+O(z^{2}),
	\end{equation}
\\
(e)As $z\to\xi$ along any ray $\xi+e^{i\phi}\Re_{+}$ with $|\phi|\leq c \leq\pi$\cite{[8]}
\begin{equation}\label{eq:3.5}
|T(z,\xi)-T_{0}(\xi)(z-\xi)^{i\kappa(\xi)}|\leq C\parallel r\parallel_{H^{1}}(\Re)|z-\xi|^{1/2},
\end{equation}
where $T_{0}(\xi)$ is the complex unit
\\
$$T_{0}(\xi)=\operatorname*{\prod}\limits_{k\in\Delta_{\xi}^{-}}\Bigg(\frac{\xi-z_{k}^{*}}{\xi-z_{k}}\Bigg)e^{i\beta(\xi,\xi)}=exp\Bigg[i\Bigg(\beta(\xi,\xi)-2\operatorname*{\sum}\limits_{k\in\Delta_{\xi}^{-}}arg(\xi-z_{k})\Bigg)\Bigg],$$
$$\beta(z,\xi)=-\kappa(\xi)log(z-\xi+1)+\int_{-\infty}^{\xi}\frac{\kappa(s)-\chi(s)\kappa(\xi)}{s-z}ds,$$
and $\chi(s)$ is the characteristic function of the interval $(\xi-1,\xi)$ and the logarithm is principally branched along $(-\infty,\xi-1]$
	\\
	\noindent\textbf{\expandafter {Proof.}}For parts (a)-(d) we can proof them directly using the definition and the Sokhotski-Plemelj formula\cite{[6]}\cite{[7]}.For part (e) we write
	\begin{equation}\label{eq:3.6}
	\begin{split}
	T(z,\xi)=\prod_{k\in\Delta_\xi^-}\Bigg(\frac{z-z_k^*}{z-z_k}\Bigg)exp\Bigg(i\int_{\xi-1}^\xi\frac{\kappa(\xi)}{s-z}ds+i\int_{-\infty}^\xi\frac{\kappa(s)-\chi(s)\kappa(\xi)}{s-z}ds\Bigg)
	\\
	=\prod_{k\in\Delta_\xi^-}\Bigg(\frac{z-z_k^*}{z-z_k}\Bigg)(z-\xi)^{i\kappa(\xi)}exp(i\beta(z,\xi)),
	\end{split}
	\end{equation}
	The result then follows from the facts that
	\begin{equation}\label{eq:3.7}
	|(z-\xi)^{i\kappa(\xi)}|\leq e^{-\pi\kappa(\xi)}=\sqrt{1+|r(\xi)|^2},
	\end{equation}
	and using Lemma 23.3 of \cite{[10]}
	\begin{equation}\label{eq:3.8}
	|\beta(z,\xi)-\beta(\xi,\xi)|\leq C||r||_{H^1(\Re)}|z-\xi|^{1/2},
	\end{equation}
	define a new  function $M^{(1)}$
	\begin{equation}\label{eq:3.9}
	M^{(1)}(z)=M(z)T(z)^{-\sigma_{3}},
	\end{equation}
	we can proof the function $M^{(1)}$ satisfies the Riemann-Hilbert problem 3.1.
\\
\noindent\textbf{\expandafter{Riemann-Hilbert problem 3.1} }Find an analysis function
$M^{(1)}:C\setminus(\Re\cup\mathcal{Z}\cup\mathcal{Z}^{*})\to SL_{2}(\mathbb{C})$ with the following properties:
\begin{enumerate}
\item $M^{(1)}(z)=I+O(z^{-1})\quad as\quad z \to\infty $
\item For each $z\in\Re$ ,the boundary values$M_{\pm}^{(1)}(z)$satisfy the jump relationship  $M_{+}^{(1)}(z)=M_{-}^{(1)}(z)J^{(1)}(z)$ where
\begin{equation}\label{eq:3.10}
J^{(1)}(z)=
\begin{cases}
\begin{pmatrix}
1&r^{*}(z)T(z)^{2}e^{-2it\theta}\\
0&1
\end{pmatrix}
\begin{pmatrix}
1&0\\
r(z)T(z)^{-2}e^{2it\theta}&1
\end{pmatrix} &{z\in(\xi,\infty)}\\
\\
\begin{pmatrix}
1&0\\
\frac{r(z)T_{-}(z)^{-2}}{1+|r(z)|^{2}|}e^{2it\theta}&1
\end{pmatrix}
\begin{pmatrix}
1&\frac{r^{*}(z)T_{+}(z)^{2}}{1+|r(z)|^{2}|}e^{-2it\theta}\\
0&1
\end{pmatrix} &{z\in(-\infty,\xi)}\\
\end{cases},
\end{equation}
\item
$M^{(1)}(z) $has simple poles at each $ z_{k} \in\Re$ and $z_{k}^{*}\in\Re^{*}$
at  which
\begin{equation}\label{eq:3.11}
\begin{split}
\operatorname*{Res}\limits_{z_{k}}M^{(1)}=
\begin{cases}
\operatorname*{lim}\limits_{z\to z_{k}}
M^{(1)}\begin{pmatrix}
0&c_{k}^{-1}(\frac{1}{T})'(z_{k})^{-2}e^{-2it\theta}\\
0&0
\end{pmatrix}&{k\in\Delta_{\xi}^{-}}\\
		\\
\operatorname*{lim}\limits_{z\to z_{k}}
M^{(1)}\begin{pmatrix}
0&0\\
c_{k}^{-1}T(z_{k})^{-2}e^{2it\theta}&0
\end{pmatrix}&{k\in\Delta_{\xi}^{+}}\\
\end{cases},
		\\
\operatorname*{Res}\limits_{z_{k}^{*}}M^{(1)}=
\begin{cases}
\operatorname*{lim}\limits_{z\to z_{k}^{*}}
M^{(1)}\begin{pmatrix}
0&0\\
-(c_{k}^{*})^{-1}T'(z_{k}^{*})^{-2}e^{-2it\theta}&0
\end{pmatrix}&{k\in\Delta_{\xi}^{-}}\\
		\\
\operatorname*{lim}\limits_{z\to z_{k}^{*}}
M^{(1)}\begin{pmatrix}
0&-c_{k}^{*}T(z_{k}^{*})^{2}e^{-2it\theta}\\
0&0
\end{pmatrix}&{k\in\Delta_{\xi}^{+}}\\
\end{cases},
\end{split}
\end{equation}
\end{enumerate}
\noindent\textbf{\expandafter {Proof.}}From above definition,we can get that $M^{(1)}$  is unimodular,analytic in $\mathbb{C}\setminus(\Re\cup\mathcal{Z}\cup\mathcal{Z}^*)$, and approaches identity as $z\to\infty$  and factorize jump (\ref{eq:3.10}) as following
\begin{equation}\label{eq:3.12}
	J^{(1)}(z)=
	\begin{cases}
	T(z)^{\sigma_s}
	\begin{pmatrix}
	1&r^{*}(z)e^{-2it\theta}\\
	0&1
	\end{pmatrix}
	\begin{pmatrix}
	1&0\\
	r(z)e^{2it\theta}&1
	\end{pmatrix}T(z)^{-\sigma_3} &{z\in(\xi,\infty)}\\
	\\
	T_-(z)^{\sigma_3}
	\begin{pmatrix}
	1&0\\
	\frac{r(z)}{1+|r(z)|^{2}|}e^{2it\theta}&1
	\end{pmatrix}
	\Bigg(\frac{T_+(z)}{T_-(z)}\Bigg)^{\sigma_3}
	\begin{pmatrix}
	1&\frac{r^{*}(z)}{1+|r(z)|^{2}|}e^{-2it\theta}\\
	0&1
	\end{pmatrix} &{z\in(-\infty,\xi)}\\
	\end{cases},
	\end{equation}
	For $k\in\Delta_\xi^+$$T(z)$ has zero at $z_k^*$ and a pole at $z_k$,so that $M_1^{(1)}=M_1(z)T(z)^{-1}$ has a removable singularity at $z_k$ and  a pole at$z_k^*$.For $M_2^{(1)}$ the situation is reversed; we have
	\begin{equation}
	\begin{split}
	M_1^{(1)}(z_k)=\operatorname*{lim}\limits_{z\to z_{k}}M_1(z)T(z)^{-1}=\operatorname*{Res}\limits_{z_k}M_1(z)(1/T)\prime(z_k)=c_ke^{2it\theta_k}M_2(z_k)(1/T)\prime(z_k)
	\\
	\operatorname*{Res}\limits_{z_{k}}M_2^{(1)}(z)=\operatorname*{Res}\limits_{z=z_k}M_2(z_k)T(z)=M_2(z_k)[(1/T)^\prime(z_k)]^{-1}=c_k^{-1}[(1/T)^\prime(z_k)]^{-2}e^{-2it\theta}M_1^{(1)}(z_k)
	\end{split},
	\end{equation}
	from which the first formula in (\ref{eq:3.11}) clearly follows.The computation of the residue at $z_k^*$ for $k\in\Delta_\xi^-$ is similar.
	\\
	\section{\noindent\textbf{\expandafter{ Introducing $\overline\partial $ extensions of jump factorization }\\\hspace*{\parindent}\vspace{2mm}}}
$\qquad $In these section,our work is to extend the  jump matrix off the real axis to new contours whose factors satisfies continuous, decaying but not analytic,we define a unknown non-analytic transformation increases nonzero $\overline\partial $ derivatives insides the regions.
\\
Define the new contours
\begin{equation}\label{eq:4.1}
\sum_{k}=\xi+e^{i(2k-1)\pi/4}\Re_{+},\quad k=1,2,3,4,
\end{equation}
Additionally,let
\begin{equation}\label{eq:4.2}
\Sigma_{R}=\Re\cup\Sigma_{1}\cup\Sigma_{2}\cup\Sigma_{3}\cup\Sigma_{4},
\end{equation}
\begin{equation}\label{eq:4.3}
\rho=\frac{1}{2}\operatorname*{min}\limits_{
\lambda,\mu\in\mathcal{Z}\cup\mathcal{Z}\cup\mathcal{Z}^{*}\lambda\not=\mu}
|\lambda-\mu|,
\end{equation}

According to our assumption, there is no pole lies on the real axis and all poles are in conjugate pairs. we have $\rho\leq dist(\mathcal{Z},\Re)$,define  $\chi_{\mathbb{Z}}\in C_{0}^{\infty}(\mathbb{C},[0,1])$ is characteristic function:
\begin{equation}\label{eq:4.4}
\chi_{\mathcal{Z}}(z)=
\begin{cases}
1&dist(z,\mathcal{Z}\cup\mathcal{Z}^{*})<\rho/3\\
0&dist(z,\mathcal{Z}\cup\mathcal{Z}^{*})>2\rho/3\\
\end{cases},
\end{equation}
	
$$\begin{tikzpicture}[node distance=1cm]
	\draw[->](-5,0)--(5,0) node[right] {$Rez$};
	\draw(0,0)--(3,3) node[below]{$\Sigma_1$};
	\draw(0,0)--(-3,3) node[below]{$\Sigma_2$};
	\draw(0,0)--(-3,-3) node[above]{$\Sigma_3$};
	\draw(0,0)--(3,-3) node[above]{$\Sigma_4$};
	\draw(0,0)node[below] {$\xi$};
	\draw(0.5,0.3) node[right]{$\Omega_1$};
	\draw(0,0.5) node[above]{$\Omega_2$};
	\draw(-0.5,0.3) node[left]{$\Omega_3$};
	\draw(-0.5,-0.3) node[left]{$\Omega_4$};
	\draw(0,-0.5) node[below]{$\Omega_5$};
	\draw(0.5,-0.3) node[right]{$\Omega_6$};
	\draw(2,1) node[right]{$\mathcal{R}^{(2)}==\begin{pmatrix}
		1&0\\
		-R_1e^{2it\theta}&1
		\end{pmatrix}$};
	\draw(0,3) node[below]{$\mathcal{R}^{(2)}==\begin{pmatrix}
		1&0\\
		0&1
		\end{pmatrix}$};
	\draw(0,-3) node[above]{$\mathcal{R}^{(2)}==\begin{pmatrix}
		1&0\\
		0&1
		\end{pmatrix}$};
	\draw(-2,-1) node[left]{$\mathcal{R}^{(2)}==\begin{pmatrix}
		1&0\\
		-R_4e^{2it\theta}&1
		\end{pmatrix}$};
	\draw(-2,1) node[left]{$\mathcal{R}^{(2)}==\begin{pmatrix}
		1&-R_3e^{2it\theta}\\
		0&1
		\end{pmatrix}$};
	\draw(2,-1) node[right]{$\mathcal{R}^{(2)}==\begin{pmatrix}
		1&-R_6e^{2it\theta}\\
		0&1
		\end{pmatrix}$};
\end{tikzpicture}$$

\noindent\textbf{\expandafter {Lemma 4.1}} Define function $R_{j}\to\mathbb{C}.J=1,3,4,6,$with boundary values satisfying
\begin{equation}\label{eq:4.5}
R_{1}(z)=
\begin{cases}
r(z)T(z)^{-2}&z\in(\xi,\infty)\\
r(\xi)T_{0}(\xi)^{-2}(z-\xi)^{-2i\kappa(\xi)}(1-\chi_{\mathbb{Z}}(z))&z\in\Sigma_{1}
\end{cases},
\end{equation}
\\
\begin{equation}\label{eq:4.6}
R_{3}(z)=
\begin{cases}
\frac{r(z)^{*}}{1+|r(z)|^{2}}T_{+}(z)^{2}&z\in(-\infty,\xi)\\
\frac{r(\xi)^{*}}{1+|r(\xi)|^{2}}T_{0}^{2}(\xi)^{2}(z-\xi)^{2i\kappa(\xi)}(1-\chi_{\mathbb{Z}}(z))
&z\in\Sigma_{2}
\end{cases},
\end{equation}
\\
\begin{equation}\label{eq:4.7}
R_{4}(z)=
\begin{cases}
\frac{r(z)}{1+|r(z)|^{2}}T_{-}(z)^{-2}&z\in(-\infty,\xi)\\
\frac{r(\xi)}{1+|r(\xi)|^{2}}T_{0}^{-2}(\xi)^{2}(z-\xi)^{-2i\kappa(\xi)}(1-\chi_{\mathbb{Z}}(z))
&z\in\Sigma_{3}
\end{cases},
\end{equation}
\\
\begin{equation}\label{eq:4.8}
R_{6}(z)=
\begin{cases}
r(z)^{*}T(z)^{2}&z\in(\xi,\infty)\\
r(\xi)^{*}T_{0}(\xi)^{2}(z-\xi)^{2i\kappa(\xi)}(1-\chi_{\mathbb{Z}}(z))&z\in\Sigma_{4}
\end{cases},
\end{equation}
such that for a fixed constant $c_{1}=c_{1}(q_{0})$,and a characteristic  function $\chi_{\mathcal{Z}}\in\mathbb{C}_{0}^{\infty}(\mathbb{C},[0,1])$ satisfying  (\ref{eq:4.4}) we have
\begin{equation}\label{eq:4.9}
\begin{split}
|R_{j}(z)|\leq c_{1}sin^{2}(arg(z-\xi))+c_{1}<Re z>^{-1/2},
\\
|\overline\partial R_{j}(z)|\leq c_{1}\overline\partial\chi_{\mathbb{Z}}(z)+c_{1}|r'(Re z)|+c_{1}|z-\xi|^{-1/2},
\\
\overline\partial R_{j}(z)=0\quad if \quad dist(z,\mathcal{Z}\cup\mathcal{Z}^{*})\leqslant\rho/3,
\end{split}
\end{equation}
Moreover,if we set $R:(\mathbb{C\setminus\sum_\Re)}\to\mathbb{C}$ by $\Re(z)|_{z\in\Omega_{j}}=\Re_{j}(z)$,$(with \Re_{2}(z)=\Re_{5}(z)=0)$the extension can be made such that $ \overline{\Re(\overline{z})}=\Re(z) $.

Next we construct the RHP $M^{(2)}$ which continuous to the real axis   and   deform its  jump matrix into the  $\Sigma_{k}$ ,let
\begin{equation}\label{eq:4.10}
M^{(2)}(z)=M^{(1)}(z)\mathcal{R}^{(2)}(z),
\end{equation}
\begin{equation}\label{eq:4.11}
\mathcal{R}^{(2)}(z)=
\begin{cases}
\begin{pmatrix}
1&0\\
-R_{1}(z)e^{2it\theta}&1
\end{pmatrix}&{z\in\Omega_{1}},\\
\\
\begin{pmatrix}
1&-R_{3}(z)e^{-2it\theta}\\
0&1
\end{pmatrix} &{z\in\Omega_{3}},\\
\\
\begin{pmatrix}
1&0\\
-R_{4}(z)e^{2it\theta}&1
\end{pmatrix}&{z\in\Omega_{4}} ,\\
\\
\begin{pmatrix}
1&-R_{6}(z)e^{-2it\theta}\\
0&1
\end{pmatrix}&{z\in\Omega_{6}},\\
\\
\begin{pmatrix}
1&0\\
0&1
\end{pmatrix}&{z\in\Omega_{2}\cup\Omega_{5}},
\end{cases}
\end{equation}
let $\Sigma^{(2)}=\cup_{j=1}^{4}\Sigma_{k}$ and the
$M^{(2)}$ satisfies the following $\overline\partial$-Riemann-Hilbert problem.

\noindent\textbf{\expandafter{$\overline\partial $-Riemann-Hilbert problem 4.1}}Find a function $M^{(2)}:\mathbb{C}\setminus(\Sigma^{(2)}\cup\mathcal{Z}\cup\mathcal{Z}^*)\to SL_{2}(\mathbb{C})$ with the following properties.
\begin{enumerate}
	\item $M^{(2)}$ is continuous and its first derivatives is sectionally continuous in $\mathbb{C}\setminus(\Sigma^{(2)}\cup\mathcal{Z}\cup\mathcal{Z}^{*})$
	\item$M^{(2)}(z)=I+O(z^{-1})\quad as\quad z \to\infty $
	\item For each $z\in\sum^{(2)}$ ,the boundary values satisfy the jump relationship  $M_{+}^{(2)}(z)=M_{-}^{(2)}(z)J^{(2)}(z)$ where
	$$J^{(2)}(z)=I+(1-\chi_{\Re}(z))\delta J^{(2)},$$
	\begin{equation}\label{eq:4.12}
	\delta J^{(2)}(z)=
	\begin{cases}
	\begin{pmatrix}
	0&0\\
	r(\xi){T_{0}(\xi)}^{-2}(z-\xi)^{-2i\kappa(\xi)}e^{2it\theta}&0
	\end{pmatrix}&{z\in\Sigma_{1}},\\
	\\
	\begin{pmatrix}
	0&\frac{r(\xi)^{*}T_{0}(\xi)^{2}}{1+|r(\xi)|^{2}}(z-\xi)^{2i\kappa(\xi)}e^{-2it\theta}\\
	0&0
	\end{pmatrix} &{z\in\Sigma_{2}},\\
	\\
	\begin{pmatrix}
	0&0\\
	\frac{r(\xi)T_{0}(\xi)^{-2}}{1+|r(\xi)|^{2}}(z-\xi)^{-2i\kappa(\xi)}e^{2it\theta}&0
	\end{pmatrix}&{z\in\Sigma_{3}},\\
	\\
	\begin{pmatrix}
	0&r(\xi)^{*}{T_{0}(\xi)}^{2}(z-\xi)^{2i\kappa(\xi)}e^{-2it\theta}\\
	0&0
	\end{pmatrix} &{z\in\Sigma_{4}},\\
	\end{cases}
	\end{equation}
	\item For $\mathbb{C}\setminus(\Sigma^{(2)}\bigcup\mathcal{Z}\bigcup\mathcal{Z}^{*})$ we have $\overline\partial M^{(2)}=M^{(2)}\overline\partial\mathcal{R}^{(2)}(z)$ where
	\begin{equation}\label{eq:4.13}
	\overline\partial \mathcal{R}^{(2)}(z)=
	\begin{cases}
	\begin{pmatrix}
	0&0\\
	\overline\partial R_{1}(z)e^{2it\theta}&0
	\end{pmatrix}&{z\in\Omega_{1}},\\
	\\
	\begin{pmatrix}
	0&\overline\partial R_{3}(z)e^{-2it\theta}\\
	0&0
	\end{pmatrix} &{z\in\Omega_{3}},\\
	\\
	\begin{pmatrix}
	0&0\\
	\overline\partial R_{4}(z)e^{2it\theta}&0
	\end{pmatrix}&{z\in\Omega_{4}},\\
	\\
	\begin{pmatrix}
	0&\overline\partial R_{6}(z)e^{-2it\theta}\\
	0&0
	\end{pmatrix} &{z\in\Omega_{6}} ,\\
	$$ 0$$&{elsewhere},\\
	\end{cases}
	\end{equation}
	\item $M^{(2)}$ has simple poles at each $z_{k}\in\Re$ and $z_{k}^{*}\in\Re^{*}$ at which
	\begin{equation}\label{eq:4.14}
	\begin{split}
	\operatorname*{Res}\limits_{z_{k}}M^{(2)}=
	\begin{cases}
	\operatorname*{lim}\limits_{z\to z_{k}}
	M^{(2)}\begin{pmatrix}
	0&c_{k}^{-1}(\frac{1}{T})'(z_{k})^{-2}e^{-2it\theta}\\
	0&0
	\end{pmatrix}&{k\in\Delta_{\xi}^{-}}\\
	\\
	\operatorname*{lim}\limits_{z\to z_{k}}
	M^{(2)}\begin{pmatrix}
	0&0\\
	c_{k}^{-1}T(z_{k})^{-2}e^{2it\theta}&0
	\end{pmatrix}&{k\in\Delta_{\xi}^{+}}\\
	\end{cases},
	\\
	\operatorname*{Res}\limits_{z_{k}^{*}}M^{(2)}=
	\begin{cases},
	\operatorname*{lim}\limits_{z\to z_{k}^{*}}
	M^{(2)}\begin{pmatrix}
	0&0\\
	-(c_{k}^{*})^{-1}T'(z_{k}^{*})^{-2}e^{-2it\theta}&0
	\end{pmatrix}&{k\in\Delta_{\xi}^{-}}\\
	\\
	\operatorname*{lim}\limits_{z\to z_{k}^{*}}
	M^{(2)}\begin{pmatrix}
	0&-c_{k}^{*}T(z_{k}^{*})^{2}e^{-2it\theta}\\
	0&0
	\end{pmatrix}&{k\in\Delta_{\xi}^{+}}\\
	\end{cases},
	\end{split}
	\end{equation}
\end{enumerate}

\noindent\textbf{\expandafter {Remark4.1}}Considering  the $\overline\partial-$RHP for $M^{(2)}$ above,though (\ref{eq:4.13}) suggest that  $M^{(2)}$ is non-analytic near the small neighborhoods at each point of discrete spectrum ,we regard  $M^{(2)}$ is analytic in $\mathbb{C}$ as its $\overline\partial$-derivative vanishes in small neighborhoods of the each point of the discrete spectrum.And we also get its jump matrices approach identity point-wise,The final two sections construct the solution $M^{(2)}$ as follows:
\begin{enumerate}
	\item We prove  the existence of the solution of the pure  Riemann-Hilbert problem which the $\overline\partial$ component of nonanalytic $\overline\partial$-RHP4.1 is ignored and compute its asymptotic expansion.
	\item Considering the existence of the solution of the  $\overline\partial$ problem and prove its solution are bound.
\end{enumerate}
\section{\noindent\textbf{\expandafter{Removing the Riemann-Hilbert component of the solution and analysis of the remaining $\overline\partial$-problem }\\\hspace*{\parindent}\vspace{2mm}}}
$\quad$At this section, we define  $M_{RHP}^{(2)} $ as  the pure Riemann-Hilbert problem of  $\overline\partial $-RHP4.1 when  $\overline\partial R^{(2)}\equiv 0 $,we will prove the solution of  $M_{RHP}^{(2)}$ exists and construct its asymptotic expansion for large $t$,and we will prove when  reduce $M_{RHP}^{(2)} $ the  $\overline\partial $-RHP-4.1 become  a pure $\overline\partial $ problem.

\noindent\textbf{\expandafter {proposition5.1}}Suppose that $M_{RHP}^{(2)}$ is a solution of pure Riemann-Hilbert problem.Define
a continuously differentiable function
\begin{equation}\label{eq:5.1}
M^{(3)}(z):=M^{(2)}(z)M_{RHP}^{(2)}(z)^{-1},
\end{equation}
satisfying the following $\overline\partial$-problem.

\noindent\textbf{\expandafter {$\overline\partial$ Problem 5.1}}Find a function $M^{(3)}:\mathbb{C}\to SL_{2}(\mathbb{C})$ with the following properties,
\begin{enumerate}
\item $M^{(3)}$ is continuous and its first derivatives is sectionally continuous in
	$\mathbb{C}\setminus\Re\cup\Sigma^{(2)}).$
	\item $M^{(3)}=I+O(z^{-1})$
	\item For$ z\in\mathbb{C}$,we have
	\begin{equation}\label{eq:5.2}
	\overline\partial M^{(3)}=M^{(3)}(z)W^{(3)},
	\end{equation}
\end{enumerate}
where $W^{(3)}:=M_{RHP}^{(2)}(z)\overline\partial R^{2}(z){M_{RHP}^{(2)}}^{-1}$ and $\overline\partial R^{(2)}$ is defined by above.
\\
\noindent\textbf{\expandafter {Proof.}}

From the (\ref{eq:5.1}) we know that the properties of the  $M^{(3)}$ is inherits from $M^{(2)}$  and $M_{RHP}^{(2)}$,both of them are continuous differentiability in $\mathbb{C}\setminus\Sigma^{(2)}$,unimodular and approach identity as $z\to\infty$,according to jump relationship
\begin{equation}\label{eq:5.3}
\begin{split}
{M^{(3)}_-}^{-1}M_+^{(3)}=M_{(RHP-)}^{(2)}(z)M_-^{(2)}(z)^{-1}M_+^{(2)}(z)M_{RHP+}^{(2)}(z)^{-1}
\\
=M_{(RHP-)}^{(2)}(z)J^{(2)}(z)(M_{RHP-}^{(2)}(z)J^{(2)}(z))^{-1}=I,
\end{split}
\end{equation}
As both the $M^{(2)}$ and $M_{RHP}^{(2)}$ can regard as analytic function when deleted neighborhood of each point of discrete spectrum $z_k$ and satisfy the residue relation (\ref{eq:4.14}),we denote  constant nilpotent matrix $N_k$ then get the Laurent expansions
\begin{equation}\label{eq:5.4}
\begin{split}
M^{(2)}(z)=C_0\Bigg[\frac{N_k}{z-z_k}+I\Bigg]+\mathcal{O}(z-z_k),
\\
M_{RHP}^{(2)}(z)^{-1}=\Bigg[\frac{-N_k}{z-z_k}+I\Bigg]\hat{C}_0+\mathcal{O}(z-z_k),
\end{split}
\end{equation}
where $C_0$ and $\hat{C}_0$ are the constant terms,this implies that
\begin{equation}\label{eq:5.5}
M^{(2)}(z)M_{RHP}^{(2)}(z)^{-1}=\mathcal{O}(1),
\end{equation}
we know that the $M^{(3)}$ has only removable singularities at each $z_k$
\begin{equation}\label{eq:5.6}
\overline\partial M^{(3)}(z)=\overline\partial M^{(2)}(z)M_{RHP}^{(2)}(z)^{-1}=M^{(2)}\overline\partial\mathcal{R}^{(2)}M_{RHP}^{(2)}(z)^{-1}=M^{(3)}W^{(3)}(z),
\end{equation}
The exist of the $M^{(3)}(z)$ prove in the next section,so
$\overline\partial$-Problem 5.1 is equivalent to the integral equation
\begin{equation}\label{eq:5.7}
M^{(3)}(z)=I-\frac{1}{\pi}\iint_{\mathbb{C}}\frac{\overline\partial M^{(3)}(s)}{s-z}\mathrm{d}A(s)=I-\frac{1}{\pi}\iint_{\mathbb{C}}\frac{M^{(3)}(s)W^{(3)}(s)}{s-z}\mathrm{d}A(s),\end{equation}
where $\mathrm{d}A(s)$ is Lebesgue measure.
\\
Using operator notation the equation (\ref{eq:5.7}) can be written as
\begin{equation}\label{eq:5.8}
(1-S)[M^{(3)}(z)]=I,
\end{equation}
where $S$ is the solid Cauchy operator
\begin{equation}\label{eq:5.9}
S[f](z)=-\frac{1}{\pi}\iint_{\mathbb{C}}\frac{f(s)W^{(3)}(s)}{s-z}\mathrm{d}A(s),
\end{equation}
In the next we will show that when $t$ is sufficiently large,the $S$ is small-norm operator,so $(1-S)^{-1}$ exists and can be expressed as a Neumann series.
\\
\noindent\textbf{\expandafter {Proposition 5.1.}} There exists a constant $C$ such that for all $t>0$ ,the operator(\ref{eq:55.9}) satisfies the inequality
\begin{equation}\label{eq:5.10}
||S||_{L^{\infty}\to L^{\infty}}\leqslant Ct^{-1/4},
\end{equation}
\noindent\textbf{\expandafter {Proof.}}We only discuss the matrix function in the region
$\Omega_1$,Let $A\in L^{\infty}(\Omega_1)$ and $s=u+iv$,
\begin{equation}\label{eq:5.11}
\begin{split}
|S[A](z)|\leqslant\iint_{\Omega_1}\frac{|A(s)M_{RHP}^{(2)}(s)W^{(2)}(s)M_{RHP}^{(2)}(s)^{-1}|}{|s-z|}dA(s),
\\
\leqslant||A||_{L^{\infty}(\Omega_1)}||M_{RHP}^{(2)}||_{L^{\infty}(\Omega_{1}^\prime)}
\iint_{\Omega_1}\frac{|\overline\partial R_1(s)|e^{-4tv(u-\xi)|}}{|s-z|}dA(s),
\end{split}
\end{equation}
Where $\Omega_1^\prime:=\Omega_1\cap(1-\chi_\mathcal{Z})$ is bounded away from the poles $z_k$ of $M_{RHP}^{(2)}$,so that $||(M_{RHP}^{(2)})^{\pm1}||_{L^\infty(\Omega_1^\prime)}$ are finite.
Using the append B we get :
\begin{equation}\label{eq:5.12}
||S||_{L^\infty\to L^\infty}\leqslant C(I_1+I_2+I_3)\leqslant Ct^{-1/4},
\end{equation}
where
\begin{equation}\label{eq:5.13}
\begin{split}
I_1=\iint_{\Omega_1}\frac{|\chi_\mathcal{Z}(s)|e^{-4tv(u-\xi)}}{|s-z|}dA(s),
\\
I_2=\iint_{\Omega_1}\frac{|r^\prime(u)||e^{-4tv(u-\xi)}}{|s-z|}dA(s),
\\
I_3=\iint_{\Omega_1}\frac{|s-\xi|^{-1/2}e^{-4tv(u-\xi)}}{|s-z|}dA(s),
\end{split}
\end{equation}

Giving the $z^{-1}$ in the laurent expansion of $M^{(3)}$  at infinity to consider the asymptotic behavior of the $q(x,t)$
\begin{equation}\label{eq:5.14}
M^{(3)}=I-\frac{1}{\pi}\iint_{\mathbb{C}}\frac{M^{(3)}(s)W^{(3)}(s)}{s-z}dA(s)=I+\frac{M_1}{z}+\frac{1}{\pi}\iint_{\mathbb{C}}\frac{sM^{(3)}(s)W^{(3)}(s)}{z(s-z)}dA(s),
\end{equation}
where
\begin{equation}\label{eq:5.15}
M_1^{(3)}=\frac{1}{\pi}\iint_{\mathbb{C}}M^{(3)}(s)W^{(3)}dA(s),
\end{equation}
\noindent\textbf{\expandafter {Proposition 5.2.}}  For all $t>0$ there exists a constant $c$ such that
\begin{equation}\label{eq:5.16}
|M_{1}^{(3)}\leqslant ct^{-3/4},
\end{equation}

proof of Proposition 5.2 is detailed in Appendix B.

\section{\noindent\textbf{\expandafter{Analysis and prove the exist the pure Riemann-Hilbert problem}\\\hspace*{\parindent}\vspace{2mm}}}
\emph{6.1. Constructing the model  problems}
\\
In this section,recall the definition of the $\rho$ ,we restrict the N-soliton in the $\mathbb{C}\setminus U_{\xi}$
\begin{equation}\label{eq:6.1}
U_{\xi}=\{z:|z-\xi|<\rho/2\},
\end{equation}
and construct solution of the form:
\begin{equation}\label{eq:6.2}
M_{RHP}^{(2)}(z)=
\begin{cases}
E(z)M^{(out)}(z)&{|z-\xi|>\rho/2},\\
E(z)M^{(\xi)}(z)&{|z-\xi|<\rho/2},
\end{cases}
\end{equation}
where  $M^{(out)}$ is the RHP which only concerning the N-soliton,the error $E(z)$ as a small-norm Riemann-Hilbert problem.  $M^{(\xi)}$ concerning the jump relation between   $M^{(out)}(z)$ and  $M^{(2)}_{RHP}(z)$
\\

\emph{6.1.1. The outer model:an N-soliton potential}

The matrix $M_{RHP}^{(2)}$ is pure-RHP ,it is  meromorphic away from the contour $\Sigma^{(2)}$, and its boundary values satisfy the jump relation (\ref{eq:4.12}) on the  $\Sigma^{(2)}$,and jump is uniformly near identity at any distance from the $\xi$,and we get the norm
\begin{equation}\label{eq:6.3}
||J^{(2)}-I||_{L^{\infty}(\Sigma^{(2)})}=O(\rho^{-2}e^{-4t|z-\xi|^{2}}),
\end{equation}
show the jumps is exponentially small outside $U_{\xi}$
so we construct the model outside $U_{\xi}$ which ignores the jumps completely.
\\
\noindent\textbf{\expandafter {Riemann-Hilbert Problem 6.1.}}  Find an analytic function $M^{(out)}:(\mathbb{C}\setminus\Re\bigcup\Re^{*})\to SL_{2}(\mathbb{C})$
such that
\begin{enumerate}
	\item $M^{(out)}(z)=I+O(z^{-1})$ as $z\to\infty$
	\item$M^{(out)}$ has simple poles at each $z_{k}\in\Re$ and $z_{k}^{*}\in\Re^{*}$ satisfying the residue relations in (\ref{eq:4.13}) with $M^{(out)}(z)$ replacing $M^{(2)}(z)$.
\end{enumerate}

\noindent\textbf{\expandafter {Proposition 6.1}} There exist a unique solution $ M^{(out)} $of RHP 6.1,specifically,
\begin{equation}\label{eq:6.4}
M^{(out)}(z)=m^{\bigtriangleup_{\xi}^{-}}(z|\sigma_{d}^{out}),
\end{equation}
where $m^{\bigtriangleup_{\xi}^{-}}$ is the solution of RHP A.2 with $\bigtriangleup=\bigtriangleup_{\xi}^{-}$ and $\sigma_{d}^{(out)}:=\{z_{k},\widetilde{c}_{k}(\xi)\}_{k=1}^{N}$ with
\begin{equation}\label{eq:6.5}
\widetilde{c}_{k}(\xi)=c_{k}exp\Bigg(\frac{i}{\pi}\int_{-\infty}^{\xi}log(1+|r(s)|^{2})\frac{ds}{s-z_{k}}\Bigg),
\end{equation}
Moreover,
$$\operatorname*{lim}\limits_{z\to\infty}2izM_{12}^{(out)}(z;x,t)=q_{sol}(x,t;\sigma_{d}^{out}),$$
where $q_{sol}(x,t;\sigma_{d}^{out})$ is the $N-soliton$ solution of (\ref{eq:1.1}) corresponding to the discrete scattering data $\sigma_{d}^{(out)}$.
\\

\emph{6.1.2. Local model near the saddle point $z=\xi$}
\\

According to analysis the jump relation (\ref{eq:4.13}),it shows at any distance from the saddle point $z=\xi$, the jump is uniformly near identity, so we construct $M^{(out)}$ only consider its N solitons without any jump,considering (\ref{eq:6.3}),it shows when $z\in U_{\xi}$ the bound  gives a point-wise,but not uniform on the decay of the jump $J^{(2)}$ to identity.In order to make the jump uniformly,we introduce  the function $E(z)$.At first ,we  introduce $M^{(\xi)}$ to make $M_{RHP}^{(out)}$  matches the  jump on $\Sigma^{(2)}\cap U_{\xi}$.

In order to use the jumps of the parabolic cylinder model problem(\ref{eq:C.3}).Define  $\zeta=\zeta(z)$
\begin{equation}\label{eq:6.6}
\zeta=\zeta(z)=2\sqrt{t}(z-\xi)\quad\Rightarrow \quad 2t\theta=\zeta^{2}/2-2t\xi^{2},
\end{equation}
which maps $U_{\xi}$ to an expanding neighborhood of $\zeta=0$,Additionally,let
\begin{equation}\label{eq:6.7}
r_{\xi}=r(\xi)T_{0}(\xi)^{-2}e^{2i(\kappa(\xi)log(2\sqrt{t}-t\xi^2)},
\end{equation}
Since $1-\chi_{\Re}(z) \equiv 1$ for $z\in U_{\xi}$,the jumps of $M_{RHP}^{(2)}$ in $U_{\xi}$ can be expressed as
\begin{equation}\label{eq:6.8}
J^{(2)}\Bigg\arrowvert_{z\in U_{\xi}}=
\begin{cases}
\begin{pmatrix}
1&0\\
r_{\xi}\zeta(z)^{-2i\kappa(\xi)}e^{i\zeta(z)^{2}/2}&1
\end{pmatrix}&{z\in\Sigma_{1}},\\
\\
\begin{pmatrix}
1&\frac{r_{\xi}^{*}}{1+|r_{\xi}|^{2}}\zeta(z)^{2i\kappa(\xi)}e^{-i\zeta(z)^{2}/2}\\
0&1
\end{pmatrix} &{z\in\Sigma_{2}},\\
\\
\begin{pmatrix}
1&0\\
\frac{r_{\xi}}{1+|r_{\xi}|^{2}}\zeta(z)^{-2i\kappa(\xi)}e^{i\zeta(z)^{2}/2}&1
\end{pmatrix}&{z\in\Sigma_{3}},\\
\\
\begin{pmatrix}
1&r_{\xi}^{*}\zeta(z)^{2i\kappa(\xi)}e^{-i\zeta(z)^{2}/2}\\
0&1
\end{pmatrix} &{z\in\Sigma_{4}},\\
\end{cases}
\end{equation}
We calculate the solution in Appendix C,define the local model  $M^{(\xi)} $in (\ref{eq:6.2}) by
\begin{equation}\label{eq:6.9}
M^{\xi}(z)=M^{(out)}(z)M^{(PC)}(\zeta(z),r_{\xi}),z\in U_{\xi},
\end{equation}
That we know that $M^{(out)}$ is analytic and bounded function in $U_{\xi}$ so the $M^{(\xi)}$ inherits the  jump $J^{(2)}$ of $M_{RHP}^{(2)}$.

\emph{6.2. The small-norm Riemann-Hilbert problem for E(z)}

Recall the definition of (\ref{eq:6.2}),the unknown function $E(z)$ is analytic in  $\mathbb{C}\setminus\Sigma^{(E)}$,
\begin{equation}\label{eq:6.10}
\Sigma^{(E)}=\partial U_{\xi}\cup (\Sigma^{(2)}\backslash U_{\xi}),
\end{equation}
where we orient $\partial U_{\xi}$ in  clockwise and E(z) satisfy the following the  small norm Riemann-Hilbert problem.
\\
\noindent\textbf{\expandafter {Riemann-Hilbert Problem 6.2}} Find a holomorphic function $E:\mathbb{C}\setminus\Sigma^{(E)}\to SL_{2}(\mathbb{C})$  with the following proprieties:
\begin{enumerate}
	\item $E(z)=I+O(z^{-1})$ as $z\to\infty$,
	\item For each $z\in\Sigma^{(E)}$  the boundary values $E_{\pm}(z)$ satisfy $E_{+}(z)=E_{-}(z)J^{(E)}(z)$ where
\begin{equation}\label{eq:6.11}
	J^{(E)}=
	\begin{cases}
	M^{(out)}(z)J^{(2)}(z)M^{(out)}(z)^{-1}&{z\in\Sigma^{(2)}\setminus U_{\xi}},\\
	\\
	M^{(out)}(z)J^{(2)}(z)M^{(PC)}(\zeta(z),r_{\xi})M^{(out)}(z)^{-1}&{z\in\partial U_{\xi}},\\
	\end{cases}
	\end{equation}
\end{enumerate}
and we can find its uniformly vanishing bound on $J_{E}-I$ are
\begin{equation}\label{eq:6.12}
|J_{E}(z)-I|=
\begin{cases}
\mathcal{O}(\rho^{-2}e^{-4t|z-\xi|^{2})}&{z\in\Sigma^{(E)}\setminus U_{\xi}}\\
\mathcal{O}(t^{-1/2})&{z\in\partial U_{\xi}}
\end{cases},
\end{equation}
then
\begin{equation}\label{eq:6.13}
||<\bullet>^{k}(J_{E}-I)||_{L^{p}(\Sigma^{E})}=\mathcal{O}(t^{-1/2}) \qquad {p\in[1,+\infty],k\geqslant 0 },
\end{equation}

The  RHP-6.2 as a small-norm Riemann-Hilbert problem, it has  a well known existence and uniqueness  theorem,we may write
\begin{equation}\label{eq:6.14}
E(z)=I+\frac{1}{2\pi i}\int_{\Sigma^{E}}\frac{(I+\eta(s))(J_{E}(s)-I)}{s-z}ds,
\end{equation}
where $\eta\in\L^{2}(\Sigma^{(E)})$ is the unique solution of
\begin{equation}\label{eq:6.15}
(I-C_{J^{(E)}})\eta=C_{J^{(E)}}I,
\end{equation}
Define $C_{V^{(E)}}=:L^{2}(\Sigma^{(E)})\to L^{2}(\Sigma^{(E)} )$ by
\begin{equation}\label{eq:6.16}
C_{J^{(E)}}f=C_{-}(f(J_{E}-I)),
\end{equation}
\begin{equation}\label{eq:6.17}
C_{-}f(z)=\operatorname*{lim}\limits_{z\to\Sigma_{-}^{(E)}}\frac{1}{2\pi i}\int_{\Sigma^{(E)}}f(s)\frac{ds}{s-z},
\end{equation}
where $C_{-}$ is the Cauchy operator, and we know that
\begin{equation}\label{eq:6.18}
||C_{V^{(E)}}||_{L_{op}^{2}(\Sigma^{(E)})}\lesssim
||C_{-}||_{L_{op}^{2}(\Sigma^{(E)}}||V^{(E)}-I||_{L^{\infty}(\Sigma^{(E)}}\lesssim \mathcal{O}(t^{-1/2}),
\end{equation}
The operator $(1-C_{V^{(E)}})^{-1}$ guarantees the exist of both $\eta$ and $E$,so it is reasonable to define the
$M_{RHP}^{(2)}(z)$ given by (\ref{eq:6.2}),and we can solve the proposition 5.1 to an unknown $M^{(3)}$ which satisfies the pure $\overline\partial $-Problem 5.1.

We analyze the asymptotic behavior for  large $z$ of the solution of RHP 2.1, we construct the function $E(z)$ of the form
\begin{equation}\label{eq:6.19}
E(z)=I+z^{-1}E_{1}+\mathcal{O}(z^{-2}),
\end{equation}
where
\begin{equation}\label{eq:6.20}
E_{1}=-\frac{1}{2\pi i} \int_{\Sigma^{(E)}}(I+\eta(s))(V^{(E}-I)ds,
\end{equation}
\begin{equation}\label{eq:6.21}
E_{1}=-\frac{1}{2\pi i}\oint_{\partial U_{\xi}}(V^{E}(s)-I)ds+O(t^{-1}),
\end{equation}
\begin{equation}\label{eq:6.22}
E_{1}(x,t)=\frac{1}{2i\sqrt{t}}M^{(out)}(\xi;x,t)
\begin{pmatrix}
0&\beta_{12}(r_{\xi})\\
-\beta_{12}(r_{\xi})&0
\end{pmatrix}
M^{(out)}(\xi;x,t)^{-1}+O(t^{-1}),
\end{equation}
we have
\begin{equation}\label{eq:6.23}
\beta_{12}(r_{\xi})=\beta_{21}(r_{\xi})^{*}=\alpha(\xi,+)e^{ix^{2}/(2t)-i\kappa(\xi)log|4t|},
\end{equation}
Here
\begin{equation}\label{eq:6.24}
|\alpha(\xi,+)|^{2}=|\kappa(\xi)|,
\end{equation}
\begin{equation}\label{eq:6.25}
arg \alpha(\xi,+)=\frac{\pi}{4}+arg\Gamma (i\kappa(\xi))-arg r(\xi)-4\operatorname*{\sum}\limits_{k\in\Delta_{\xi}^{-}}arg(\xi-z_{k})-2\int_{-\infty}^{\xi}log|\xi-s|d_{s}\kappa(s),
\end{equation}

\section{\noindent\textbf{\expandafter{Long time asymptotics for focusing KE}\\\hspace*{\parindent}\vspace{2mm}}}
$\qquad$In this section, we will give the details of the proof for Theorem 1.1 as  $t\to+\infty$.

\noindent\textbf{\expandafter {Proof of Theorem 1.1.}} According to transformations,we know that the solution of (1.1) can be expressed as
\begin{equation}\label{eq:7.1}
M(z)=M^{(3)}(z)E(z)M^{(out)}(z)R^{(2)}(z)T(z)^{\sigma_{3}},\quad z\in\mathbb{C}\setminus U_{\xi},
\end{equation}
let $z\to\infty$ eventually  $z\in\Omega_{2}$ so that $R^{(2)}=I$ ,we have

\begin{equation}\label{eq:7.2}
T(z)^{\sigma_{3}}=I+\frac{T_{1}\sigma_{3}}{z}+O(z^{-2}),\quad  T_{1}=2\operatorname*{\sum}\limits_{\Delta^{-}} Imz_{k}-\int_{-\infty}^{\xi}\kappa(s)\mathrm{ds},
\end{equation}
Now
\begin{equation}\label{eq:7.3}
M=\Bigg(I+\frac{M_{1}^{(3)}}{z}+\cdots\Bigg)\Bigg(I+\frac{E_{1}}{z}+\cdots\Bigg)\Bigg(I+\frac{M_{1}^{(out)}}{z}+\cdots\Bigg)\Bigg(I+\frac{T_{1}\sigma_{3}}{z}+\cdots\Bigg),
\end{equation}
the coefficient of the $z^{-1}$ in the Laurent expansion of $M$ is
\begin{equation}\label{eq:7.4}
M_{1}=M_{1}^{(3)}+E_{1}+M_{1}^{(out)}+O(t^{-3/4}),
\end{equation}
we know that
\begin{equation}\label{eq:7.5}
q(x,t)=2i(M_{1}^{(out)})_{12}+2i(E_{1})_{12}+O(t^{-3/4}),
\end{equation}
Applying Proposition 6.1 to the first term and using (\ref{eq:6.20})- (\ref{eq:6.25})to evaluate the second term ,we have
\begin{equation}\label{eq:7.6}
q(x,t)=q_{sol}(x,t;\sigma_{d}^{out})+t^{-1/2}f^{+}(x,t)+O(t^{-3/4}),
\end{equation}

We know that $q_{sol}(x,t;\sigma_{d}^{out})$ is the solution of N-soliton generated from  Proposition 6.2, we will give the relationship concerning $q_{sol}(x,t;\sigma_{d}^{out})$ with $q_{sol}(x,t;\sigma_{d}^{+}{(\mathcal{I})})$ whose contained in the   a cone $C(x_{1},x_{2},v_{1},v_{2})$ as defined in Theorem 1.1.
Using the appendix A,we know that replace $q_{sol}(x,t;\sigma_{d}^{out})$ with $q_{sol}(x,t;\sigma_{d}^{+}{(\mathcal{I})})$ their exist exponential errors which are absorbed into the $\mathcal{O}(t^{-3/4})$ term.

\noindent\textbf{\expandafter {The long-time asymptotics}}
As $t\to\infty$,such that $|\xi|=|-\frac{x}{4t}|<M$
\begin{equation}\label{eq:7.7}
2im(x,t)=(q_{sol}(x,t;\sigma^{\pm}(I))+t^{-\frac{1}{2}}f^{\pm}(x,t)+\mathcal{O}(t^{-\frac{3}{4}})),
\end{equation}
where  $q_{sol}$ and $f^\pm$ are showed above.
In order to get the asymptotic of $q_(x,t)$,we also need to calculate $e^{-2i\int_{(-\infty,t)}^{(x,t)}\Delta}$
\\
\noindent\textbf{\expandafter {Proposition  7.1}} As $t\to \infty$
\begin{equation}\label{eq:7.8}
e^{-2i\int_{(-\infty,t)}^{(x,t)}\Delta}=e^{-2i\beta\int_{-\infty}^x|q|(x\prime,t)^2dx\prime}=e^{-8i\beta\int_{(-\infty,t)}^{(x,t)}|m(x\prime,t)^2|dx\prime}+\mathcal{O}(t^{-\frac{3}{2}}),
\end{equation}
where
\begin{equation}\label{eq:7.9}
|m(x,t)|^2=\frac{1}{2}|(q_{sol}(x,t;\sigma^{\pm}(I))+t^{-\frac{1}{2}}f^{\pm}(x,t)+\mathcal{O}(t^{-\frac{3}{4}}))|^2,
\end{equation}
\appendix
\section{Appendix :Merorphic solutions of the focusing KE Riemann-Hilbert problem}
In this section,we consider the unknown meromorphic function(with the reflection coefficient $r(z)\equiv 0$ ) only has a finite discrete spectrum,we will proof the existence and uniqueness of this problem and discuss its asymptotic behavior as $t\to\infty$.

\noindent\textbf{\expandafter {Riemann-Hilbert Problem A.1}} Given discrete data $\sigma_{d}=\{(z_{k},c_{k})\}_{k=1}^{N}\in\mathbb{C}^{+}\times\mathbb{C}_*$,let $\mathcal{Z}=\{z_{k}\}_{k=1}^N.$,Find an analytic function $m:\mathbb{C}\setminus(\mathcal{Z}\cup\mathcal{Z}^*)\to SL_{2}(\mathbb{C})$ with the following properties:
\begin{enumerate}
	\item $m(z;x,t|\sigma_{d})=I+O(z^{-1})  as z\to\infty$,
	\item Each point of $\mathcal{Z}\cup\mathcal{Z}^*$ is a simple pole of $m(z;x,t|\sigma_{d})$ They satisfy the residue conditions
	\begin{equation}\label{eq:AA.1}
	\begin{split}
	\operatorname*{Res}\limits_{z=z_{k}^*}m(z;x,t|\sigma_{d})=\operatorname*{lim}\limits_{z=z_{k}^*}m(z;x,t|\sigma_{d})\sigma_2n_{k}^*\sigma_2,
	\\
	\operatorname*{Res}\limits_{z=z_{k}}m(z;x,t|\sigma_{d})=\operatorname*{lim}\limits_{z=z_{k}}m(z;x,t|\sigma_{d})n_{k},
	\end{split}
	\end{equation}
	where $n_{k}$ is the nilpotent matrix,
	\begin{equation}\label{eq:AA.2}
	n_{k}=\begin{pmatrix}
	0&0\\
	\gamma_k(x,t)&0
	\end{pmatrix}\qquad
	\gamma_{k}(x,t):=c_kexp(2i(tz_k^2+xz_k)),
	\end{equation}
\end{enumerate}
Using the Liouville's theorem to get the uniqueness of the solution and we can proof the symmetry $m(z|\sigma_d)=\sigma_2m(z^*|\sigma_d)^*\sigma_2$.it follows that any solution of RHP A.1 has the solution of the form as followed:
\begin{equation}\label{eq:AA.3}
m(z;x,t|\sigma_{d})=I+\sum_{k=1}^{N}\frac{1}{z-z_k}
\begin{pmatrix}
\alpha_k(x,t)&0\\
\beta_k(x,t)&0
\end{pmatrix}
+\frac{1}{z-z_k^*}
\begin{pmatrix}
0&-\beta_k(x,t)^*\\
0&\alpha_k(x,t)^*
\end{pmatrix},
\end{equation}
for coefficients $\alpha_k(x,t),\beta_k(x,t)$to be determined.
\\
\noindent\textbf{\expandafter {Proposition A.1}}Given data $\sigma_d=\{(z_k,c_k)\}_{k=1}^{N}\in\mathbb{C}\times\mathbb{C}_* $such that $z_j\not=z_k$ for $j\not=k$there exists a unique solution of RHP B.1 for each $(x,t)\in\Re^2$
\\
\noindent\textbf{\expandafter {Proof:}}The proof can be found in(\cite{[15]}.

\emph{A.1 Renormalizations of the reflectionless Riemann-Hilbert problem}
\\

Define the N-soliton solutions of  RHP A.1 with $r(z)=0$,$1/a(z)$ is the transmission coefficient of the reflectionless initial data.
\begin{equation}\label{eq:A.4}
m(z;x,t|\sigma_d)=\Bigg[\frac{\phi_1^{(-)}(x,t;z)}{a(z)}|\phi_2^{(+)}(x,t;z)\Bigg]e^{i(tz^2+xz)\sigma_3}\quad a(z)=\prod_{k=1}^N\Bigg(\frac{z-z_k}{z-z_k^*}\Bigg),
\end{equation}

Let  $\Delta\subset\{1,2,\dots,N\}$ and $\triangledown=\Delta^c=\{1,\dots,N\}\setminus\Delta.$.Define
\begin{equation}\label{eq:A.5}
a_{\Delta}(z)=\prod_{k\in\Delta}\Bigg(\frac{z-z_k}{z-z_k^*}\Bigg)\quad and \quad
a_{\triangledown}(z)=\frac{a(z)}{z_{\Delta}(z)}=\prod_{k\in\triangledown}\Bigg(\frac{z-z_k}{z-z_k^*}\Bigg),
\end{equation}
The renormalization
\begin{equation}\label{eq:A.6}
m^{\Delta}(z|\sigma_d)=m(z|\sigma_d)a_{\Delta}(z)^{\sigma_3}=\Bigg[\frac{\phi_1^{(-)}(x,t;z)}{a_{\triangledown}(z)}\Bigg|\frac{\phi_2^{\pm}(x,t;z)}{a_{\Delta}(z)}\Bigg]e^{i(tz^2+xz)\sigma_3},
\end{equation}
 it obvious that by choice of $\Delta$ to split the poles between the columns,and $m^{\Delta}$ satisfies followed the  modified discrete Riemann-Hilbert problem.
 \\
\noindent\textbf{\expandafter {Riemann-Hilbert Problem A.2}} Given discrete data $\sigma_d=\{(z_k,c_k)\}_{k=1}^N\subset\mathbb{C}^+\times\mathbb{C}_*$ and $\Delta\subset\{1,\dots,N\}$ find an analytic function $m^{\Delta}:\mathbb{C}\setminus(\mathcal{Z}\cup\mathcal{Z}^*)\to SL_2(\mathbb{C})$ with the following properties:
\begin{enumerate}
	\item $m^{\Delta}(z;x,t|\sigma_d)=I+O(z^{-1}) as z\to\infty$,
	\item Each point of $\mathcal{Z}\cup\mathcal{Z}^*$ is a simple pole of $m^{\Delta}(z;x,t|\sigma_d)$,they satisfy the residue conditions
	\begin{equation}\label{eq:A.7}
	\begin{split}
	\operatorname*{Res}\limits_{z=z_k} m^{\Delta}(z;x,t|\sigma_d)=\operatorname*{lim}\limits_{z=z_k} m^{\Delta}(z;x,t|\sigma_d)n_k^{\Delta},
	\\
	\operatorname*{Res}\limits_{z=z_k^*} m^{\Delta}(z;x,t|\sigma_d)=\operatorname*{lim}\limits_{z=z_k^*} m^{\Delta}(z;x,t|\sigma_d)\sigma_2(n_k^{\Delta})^*\sigma_2,
	\end{split}
	\end{equation}
\end{enumerate}
where $n_k$ is the nilpotent matrix,
\begin{equation}\label{eq:AA.8}
n_k^{\Delta}=
\begin{cases}
\begin{pmatrix}
0&0\\
\gamma_k(x,t)a_{\Delta}(z_k)^2&0
\end{pmatrix}
\\
\\
\begin{pmatrix}
0&\gamma_k(x,t)^{-1}a_{\Delta}'(z_k)^{-2}\\
0&0
\end{pmatrix}
\end{cases}
\gamma_k(x,t):=c_kexp(2i(tz_k^2+xz_k)),
\end{equation}
and $a_\Delta$ is defined in (\ref{eq:A.5})

When the poles s $z_k\in\Re$ are distinct we know that the RHP B.2 has a unique solution because it is transformation of $m(z;x,t|\sigma_d)$,the advantage of this method we will proof above that by choosing the $\Delta$ correctly,other soliton asymptotic behavior are under better control when  $t\to\infty,-x/4t=\xi$

\emph{A.2 Long time behavior of the soliton solutions}

If there is only a single solution $\sigma_d=\{(\xi+i\eta,c_1)\}$,we know that
\begin{equation}\label{eq:A.9}
q_{sol}(x,t)=q_{sol}(x,t;\{\xi+i\eta\})=2ia_{1}e^{\Omega_{1}-\Omega_{1}^{*}}(P^{-1})e^{8i\beta\int |ae^{\Omega_{1}-\Omega_{1}^{*}}(P^{-1})^{2}|dx},
\end{equation}
When there are N-soliton$N>1$,we know that the N-soliton asymptotically separate into N single-soliton as $t\to\infty$.

Define
\begin{equation}\label{eq:A.10}
\mu=\mu(\mathcal{I})=\operatorname*{min}\limits_{z_k\in\mathcal{Z}\setminus\mathcal{Z(\mathcal{I})}}\{Im(z_k)dist(Re z_k,\mathcal{I})\},
\end{equation}

\noindent\textbf{\expandafter {Proposition A.2}}. Given discrete scattering data $\sigma_d=\{(z_k,c_k)\}_{k=1}^N\subset\mathbb{C}^+\times(\mathbb{C}\setminus\{0\})$,fix $x_1,x_2,v_1,v_2\in\Re$ with $x_1\leq x_2$ and $v_1\leq v_2$,Let  $ \mathcal{I}=[-v_2/2,-v_1/2],$Then,as $t\to\pm\infty$ with $(x,t)\in C(x_1,x_2,v_1,v_2) $ we have
\begin{equation}\label{eq:A.11}
m^{\Delta_\xi^{\mp}}(z;x,t|\sigma_d)=(I+O(e^{-4\mu|t|}))m^{\Delta_\xi^\mp(\mathcal{I})}(z;x,t|\sigma_d^\pm(\mathcal{I})),
\end{equation}
for all $z$ bounded away from $\mathcal{Z}\cup\mathcal{Z}^*$.
\\
Here $\sigma_d^\pm(\mathcal{I})$  is the scattering data for the $N(\mathcal{I})\leq N$ soliton given by
\begin{equation}\label{eq:A.12}
\sigma_d^\pm(\mathcal{I})=\{(z_k,c_k^\pm(\mathcal{I})):z_k\in\mathcal{Z}(\mathcal{I})\}\qquad c_k^\pm(\mathcal{I})=c_k\prod_{z_j\in\mathcal{Z}^\mp(\mathcal{I})}\Bigg(\frac{z_k-z_j}{z_k-z_j^*}\Bigg)^2,
\end{equation}

\noindent\textbf{\expandafter {Corollary A.3}}Let $q_{sol}(x,t;\sigma_d)$ is the N-soliton of the fKE equation (\ref{eq:1.1}) with its discrete scattering data
$\sigma_d=\{(z_k,c_k)\}_{k=1}^N\in\mathbb{C}^+\times(\mathbb{C}\setminus\{0\})$ and the $\mathcal{I},C(x_1,x_2,v_1,v_2) $ and $\sigma_d^\pm(\mathcal{I})$ be as given in Proposition A.2.then as $t\to\pm\infty$,with $(x,t)\in C(x_1,x_2,v_1,v_2)$
\begin{equation}\label{eq:A.13}
q_{sol}(x,t;\sigma_d)=q_{sol}(x,t;\sigma_d^\pm(\mathcal{I})+\mathcal{O}(e^{-4\mu t}),
\end{equation}
where $q_{sol}(x,t;\sigma_d^\pm(\mathcal{I})$ is the solution of the fKE with $N(I)$-soliton  and its scattering data is $\sigma_d^\pm(\mathcal{I})$.
\\
\noindent\textbf{\expandafter {Proof of Proposition A.2}} Observe that
\begin{equation}\label{eq:A.14}
|\gamma_k(x_0+vt,t)|=|c_k||exp[-2x_0Im(z_k)]exp[-4tIm(z_k)Re(z_k+v/2)],
\end{equation}
The choice of normalization $\Delta=\Delta_\xi^\mp$ in RHP A.2 ensures that $|t|\to\infty$ with $(x,t)\in C(x_1,x_2,v_1,v_2)$ that
\begin{equation}\label{eq:A.15}
||n_k^{\Delta_\xi^\mp}||=
\begin{cases}
\mathcal{O}(1)&z_k\in\mathcal{Z}(\mathcal{I})
\\
\mathcal{O}(exp(-4\mu|t|))&z_k\in\mathcal{Z}\setminus\mathcal{Z}(\mathcal{I})
\end{cases},
\end{equation}
This suggests that the residues with $z_k\in\mathcal{Z}\setminus\mathcal{Z}(\mathcal{I})$  contribute to the solution $m^{\Delta_\xi^\pm}$
insignificance.
 Around each $z_k\in\mathcal{Z}\setminus\mathcal{Z}(\mathcal{I})$
we trade its residue for a near identity jump by  introducing small disk$D_k$ whose radii are chosen sufficiently small that they are non-overlapping.We make the change of variables
\begin{equation}\label{eq:A.16}
m^{\Delta_\xi^\mp}(z|\sigma_d)=
\begin{cases}
\hat{m}^{\Delta_\xi^\mp}(z)(I+\frac{n_k}{z-z_k})&z\in D_k\\
\hat{m}^{\Delta_\xi^\mp}(z)(I+\frac{\sigma_2n_k^*\sigma_2}{z-z_k^*})&z\in D_k^*\\
\hat{m}^{\Delta_\xi^\mp}(z)&elsewhere
\end{cases},
\end{equation}
The new unknown $\hat{m}^{\Delta_\xi^\mp}(z)$ has jumps across each disk boundary which ,by  (\ref{eq:A.15}),satisfy
\begin{equation}\label{eq:A.17}
\hat{m}_+^{\Delta_\xi^\mp}(z)=\hat{m}_=^{\Delta_\xi^\mp}(z)\hat{v}(z)\qquad z\in\partial D_k\cup\partial D_k^*,
\end{equation}
with
\begin{equation}\label{eq:A.18}
||\hat{v}-I||=\mathcal{O}(exp(-4\mu|t|))\qquad  z\in\partial D_k\cup\partial D_k^*,
\end{equation}

The $m^{\Delta_\xi^\mp(\mathcal{I}}(z|\sigma_d^\pm(\mathcal{I})$ has the same poles as  $\hat m^{\Delta_\xi^\mp(\mathcal{I}}(z|\sigma)$ with the same residue conditions.that
\begin{equation}\label{eq:A.19}
e(z)=\hat{m}^{\Delta_\xi^\mp}(z|\sigma_d)[m^{\Delta_\xi^\mp(\mathcal{I}}(z|\sigma_d^\pm(\mathcal{I}))]^{-1},
\end{equation}
has no poles,and its jumps satisfy estimates identical to (\ref{eq:A.18}).

We show that $e(z)$ exists and that $e(z)=I+\mathcal{O}(e^{-4\mu|t|})$ for all sufficiently large $|t|$  by using the small-norm Riemann-Hilbert problems.
and from(\ref{eq:A.16}) and (\ref{eq:A.19}) that $m^{\Delta_\xi^\mp}(z;x,t|\sigma_d)=e(z)m^{\sigma_\xi^\mp(\mathcal{I}}(z;x,t|\sigma_d^\pm(\mathcal{I}))$  for $z$ outside each dist $D_k$ and $D_k^*$.The result follows immediately.

\section{Appendix:Detail of calculations for the $\overline\partial $problem}
\noindent\textbf{\expandafter {Proposition C.1.}}There exist constant $c_1,c_2, and c_3 $ such that for all $t>0$,the integrals $I_j,j=1,2,3,$ the defined by (\ref{eq:6.7})-(\ref{eq:6.8}) satisfy the bound
\begin{equation}\label{eq:B.1}
|I_j|\leq\frac{c_j}{t^{1/4}},j=1,2,3,
\end{equation}
\noindent\textbf{\expandafter {Proof.}}Our proof follows that found in\cite{[10]}.Let $s=u+iv$ and $z=\alpha+i\beta$.And  we use the elementary fact $||\frac{1}{s-z}||_{L_u^2(v+\xi,\infty})^2=(\int_{v+\xi}^\infty\frac{1}{(u-\alpha)^2+(v-\beta)^2}ds)^{1/2}\leq\int_\Re\frac{1}{u^2+(v-\beta)^2}du=\frac{\pi}{v-\beta}$ to show that

\begin{equation}\label{eq:B.2}
\begin{split}
|I_1|\leq\int_0^\infty\int_{v+\xi}^\infty\frac{|\chi_\mathcal{Z}(s)|}{s-z}e^{-4tv(u-\xi)}dudv\leq\int_0^\infty e^{-4tv^2}\int_{v+\xi}^\infty\frac{|\chi_\mathcal{Z}(s)|}{|s-z|}dudv
\\
\leq\int_0^\infty e^{-4tv^2}||\chi_{\mathcal{Z}}(s)||_{L_u^2(v+\xi,infty)}||\frac{1}{s-z}||_{L_u^2(v+\xi,infty)}dv
\\
\leq c_1\int_0^{\infty}\frac{e^{-4tv^2}}{|v-\beta|^{1/2}}dv\leq c_1t^{-1/4}\int_\Re\frac{e^{-4(w+\sqrt{t}\beta)^2}}{|w|^{1/2}}\leq c_1t^{-1/4}\int_\Re\frac{e^{-4w^2}}{|w|^{1/2}}\leq c_1t^{-1/4},
\end{split}
\end{equation}
The bonded for $I_2$ is similar to $I_1$,Recalling that $r\in H^{1,1}(\Re),$

\begin{equation}\label{eq:B.3}
|I_2|\leq\int_0^\infty e^{-4tv^2}\int_{v+\xi}^\infty\frac{|r\prime(u)|}{|s-z|}dudv\leq||r\prime(u)||_{L^2(\Re)}\int_0^\infty e^{-4tv^2}||\frac{1}{s-z}||_{L_u^q(v+\xi,\infty)}dv\leq\frac{c_2}{t^{1/4}},
\end{equation}
For $I_3$ choose $p>2$ and $q$ H$\ddot{o}$lder conjugate to $p$,then
\begin{equation}\label{eq:B.4}
\begin{split}
|I_3|\leq\int_0^\infty e^{-4tv^2}||(s-\xi)^{-1/2}||_{L_u^p(v+\xi,\infty)}||(s-z)^{-1}||_{L_u^q(v+\xi,\infty)}dv
\\
\leq c_p\int_0^\infty e^{-4tv^2}v^{1/p-1/2}|v-\beta|^{1/q-1}dv,
\end{split}
\end{equation}
To bound this last integral observe that
\begin{equation}\label{eq:B.5}
\begin{split}
\int_0^\beta e^{-tv^2}v^{1/p-1/2}(\beta-v)^{1/q-1}dv=\int_0^1 \beta^{1/2}e^{-t\beta^2w^2}w^{1/p-1/2}(1-w)^{1/q-1}dw
\\
\leq ct^{-1/4}\int_0^1\int_0^1 w^{1/p-1}(1-w)^{1/q-1}dw\leq Ct^{-1/4},
\end{split}
\end{equation}
where we've used the bound $e^{-m}\leq m^{-1/4}$ for $ m\geq 0$ to replace the exponential factor in the second integral,Finally
\begin{equation}\label{eq:B.6}
\int_\beta^\infty e^{-tv^2}v^{1/p-1/2}(v-\beta)^{1/q-1}dv \leq\int_0^\infty e^{-tw^2}w^{-1/2}dw\leq Ct^{-1/4},
\end{equation}
The result is confirmed.
\\
\noindent\textbf{\expandafter {Proposition B.2}} For all $t>0$ there exists a constant $c$ such that
\begin{equation}\label{eq:B.7}
|M_1^{(3)}|\leq ct^{-3/4},
\end{equation}
\noindent\textbf{\expandafter {Proof.}}The proof given here follows calculations that can be found in \cite{[9]}\cite{[14]}.recalling that the set $\Omega_1^\prime=\Omega_1\cup supp(1-\chi_\mathcal{Z})$ is bounded away from the poles of $M_{RHP}^{(2)}$,we have

\begin{equation}\label{eq:B.8}
\begin{split}
|M_1^{(3)}|\leq\iint_{\Omega_1}|M^{(3)}(s)M_{RHP}^{(2)}(s)M_{RHP}^{(2)}(s)^{-1}|dA
\leq\frac{1}{\pi}||M^{(3)}||_{L^\infty(\Omega)}||M_{RHP}^{(2)}||_{L^\infty}(\Omega^\prime)\iint_{\Omega}|\overline\partial R_1e^{2it\theta}|dA
\\
\leq C\Bigg(\iint_{\Omega_1}|\chi_\mathcal{Z}(s)|e^{-4tv(u-\xi)}dA+\iint_{\Omega_1}|r^\prime(u)|e^{-4tv(u-\xi)}dA+\iint_{\Omega_1}\frac{1}{|s-\xi}^{1/2}e^{-4tv(u-\xi)}dA
\\
\leq C(I_4+I_5+I_6),
\end{split}
\end{equation}
We bound $I_4$ by applying the Cauchy-Schwarz inequality
\begin{equation}\label{eq:B.9}
\begin{split}
|I_4|\leq\int_0^\infty||\chi_\mathcal{Z}||_{L_u^2(u+\xi,\infty)}(\int_v^\infty e^{-8uv}ds)^{1/2}dv
\\
\leq ct^{-1/2}\int_0^\infty\frac{e^{-4tv^2}}{\sqrt{v}}\leq\int_0^\infty\frac{e^{-4tv^2}}{\sqrt{w}}dw\leq ct^{-3/4},
\end{split}
\end{equation}
For $2<p<4$,
\begin{equation}\label{eq:B.10}
\begin{split}
|I_6|\leq c\int_0^\infty v^{1/p-1/2}(\int_v^\infty e^{-4qtuv}du)^{1/q}dv\leq ct^{-1/q}\int_o^\infty v^{1/p-3/2}e^{-4tv^2}dv
\\
\leq ct^{-3/4}\int_0^\infty w^{2/p-3/2}e^{-4w^2}dw\leq ct^{-3/4},
\end{split}
\end{equation}
where we have used the substitution $w=t^{1/2}$ and the fact that $-1<\frac{2}{p}-\frac{3}{2}<-\frac{1}{2}$.

\section{Appendix :The parabolic cylinder model problem}
This section we describe the long-time asymptotic calculations the integrable nonlinear waves of the parabolic cylinder model problem,\cite{[12]}
Define
\begin{equation}\label{eq:C.1}
\Sigma_{j}=\Bigg\{\zeta\in\mathbb{C}|arg\zeta=\frac{2j-1}{4}\pi\Bigg\},j=1,\dots,4,
\end{equation}
Fix  $r\in\mathbb{C}$ define
\begin{equation}\label{eq:C.2}
\kappa=\kappa(r):=-\frac{1}{2\pi}log(1+|r|^{2}),
\end{equation}
And we define six connected open sectors in  $\mathbb{C}\backslash(\Sigma_{PC}\cup\Re)$, the sequence of region is in a counterclockwise
$$
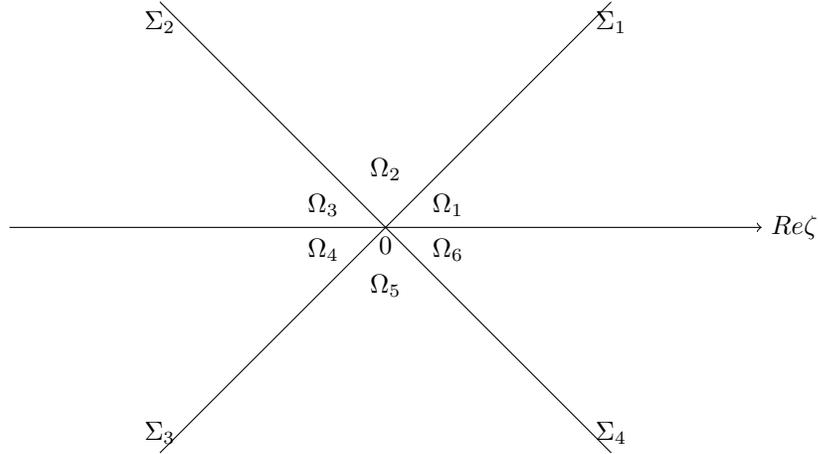
\begin{tikzpicture}[node distance=1cm]
\draw[->](-5,0)--(5,0) node[right] {$Re\zeta$};
\draw(0,0)--(3,3) node[below]{$\Sigma_1$};
\draw(0,0)--(-3,3) node[below]{$\Sigma_2$};
\draw(0,0)--(-3,-3) node[above]{$\Sigma_3$};
\draw(0,0)--(3,-3) node[above]{$\Sigma_4$};
\draw(0,0) node[below] {$0$};
\draw(0.5,0.3) node[right]{$\Omega_1$};
\draw(0,0.5) node[above]{$\Omega_2$};
\draw(-0.5,0.3) node[left]{$\Omega_3$};
\draw(-0.5,-0.3) node[left]{$\Omega_4$};
\draw(0,-0.5) node[below]{$\Omega_5$};
\draw(0.5,-0.3) node[right]{$\Omega_6$};
\end{tikzpicture}$$
\noindent\textbf{\expandafter {Parabolic Cylinder Model Riemann-Hilbert Problem A.1.}} The analytic function $M^{(PC)}(\bullet,r):\mathbb{C}\backslash\Sigma^{(PC)}\to SL_{2}(\mathbb{C})$,where $r\in\mathbb{C}$ is fixed.
\begin{enumerate}
	\item $M^{(PC)}(\zeta,r)=I+\frac{M^{(PC)}(r)}{\zeta}+O(\zeta^{-2})$,
	\item For $\zeta\in\Sigma^{(PC)}$,the continuous boundary values $M_{\pm}^{(PC)}(\zeta,r)$,satisfy the jump relation $M_{+}^{(PC)}(\zeta,r)=M_{+}^{(PC)}(\zeta,r)V^{(PC)}(\zeta,r)$,where
	\begin{equation}\label{eq:C.3}
	V^{(PC)}(\zeta,r)=
	\begin{cases}
	\begin{pmatrix}
	1&0\\
	r\zeta^{-2i\kappa}e^{i\zeta^{2}/2}&1
	\end{pmatrix}&arg\zeta=\pi/4\\
	\\
	\begin{pmatrix}
	1&r^{*}\zeta^{2i\kappa}e^{-i\zeta^{2}/2}\\
	0&1
	\end{pmatrix} &arg\zeta=-\pi/4\\
	\\
	\begin{pmatrix}
	1&0\\
	\frac{r^{*}}{1+|r|^{2}}\zeta^{2i\kappa}e^{-i\zeta^{2}/2}&1
	\end{pmatrix}&arg\zeta=3\pi/4\\
	\\
	\begin{pmatrix}
	1&0\\
	\frac{r}{1+|r|^{2}}\zeta^{-2i\kappa}e^{i\zeta^{2}/2}&1
	\end{pmatrix} &arg\zeta=-3\pi/4\\
	\end{cases},
	\end{equation}
	
	According to the solutions of the parabolic cylinder equation $(\frac{\partial^{2}}{\partial z^{2}}+(\frac{1}{2}-\frac{z^{2}}{2}+a))D_{a}(z)=0$,\cite{[8]}\cite{[14]}we have an explicit solution of the  $M^{(PC)}(\zeta,r)$.
\begin{equation}\label{eq:C.4}
	M^{(PC)}(\zeta,r)=\Phi(\zeta,r)\mathcal{P}(\zeta,r)e^{\frac{i}{4}\zeta^{2}\sigma_{3}}\zeta^{-i\kappa\sigma_{3}},
\end{equation}
	where
	\begin{equation}\label{eq:C.5}
	\mathcal{P}(\zeta,r)=
	\begin{cases}
	\begin{pmatrix}
	1&0\\
	-r&1
	\end{pmatrix}&\zeta\in\Omega_{1}\\
	\\
	\begin{pmatrix}
	1&\frac{-r^{*}}{1+|r|^{2}}\\
	0&1
	\end{pmatrix} &\zeta\in\Omega_{3}\\
	\\
	\begin{pmatrix}
	1&0\\
	\frac{r}{1+|r|^{2}}&1
	\end{pmatrix}&\zeta\in\Omega_{4}\\
	\\
	\begin{pmatrix}
	1&r^{*}\\
	0&1
	\end{pmatrix} &\zeta\in\Omega_{6}\
	\end{cases},
	\end{equation}
where
	\begin{equation}\label{eq:C.6}
	\Phi(\zeta,r)=
	\begin{cases}
	\begin{pmatrix}
	e^{-\frac{3\pi\kappa}{4}}D_{i\kappa}(e^{-\frac{3i\pi}{4}}\zeta)&-i\beta_{12}e^{\frac{\pi}{4}}D_{-i\kappa-1}(e^{-\frac{i\pi}{4}}\zeta)\\
	i\beta_{21}e^{-\frac{3\pi}{4}}D_{-i\kappa+1}(e^{-\frac{i\pi}{4}}\zeta)&e^{\frac{\pi\kappa}{4}}D_{-i\kappa}(e^{-\frac{-i\pi}{4}}\zeta)
	\end{pmatrix}&\zeta\in\mathbb{C}^{+}\\
	\\
	\begin{pmatrix}
	e^{\frac{\pi\kappa}{4}}D_{i\kappa}(e^{\frac{-i\pi}{4}}\zeta)&-i\beta_{12}e^{-\frac{3\pi}{4}(\kappa-i)}D_{-i\kappa-1}(e^{\frac{3i\pi}{4}}\zeta)\\
	i\beta_{21}e^{\frac{\pi}{4}(\kappa+i)}D_{i\kappa-1}(e^{\frac{i\pi}{4}}\zeta)&e^{-\frac{3\pi\kappa}{4}}D_{-i\kappa}(e^{\frac{3i\pi}{4}}\zeta)
	\end{pmatrix}&\zeta\in\mathbb{C}^{-}\\
	
	\end{cases},
	\end{equation}
	and $\beta_{12}$ and $\beta_{21}$ are the complex constants
	\begin{equation}\label{eq:C.7}
	\beta_{12}=\beta_{12}(r)=\frac{\sqrt{2\pi}e^{i\pi/4e^{-\pi\kappa/2}}}{r\Gamma(-i\kappa)},\qquad \beta_{21}=\beta_{21}(r)=\frac{-\sqrt{2\pi}e^{-i\pi/4e^{-\pi\kappa/2}}}{r^*\Gamma(i\kappa)},\end{equation}
	We use the  result is given in \cite{[13]},get the result as
	\begin{equation}\label{eq:C.8}
	M^{PC}(\zeta,r)=I+\frac{1}{\zeta}
	\begin{pmatrix}
	0&-i\beta_{12}(r)\\
	i\beta_{21}(r)&0
	\end{pmatrix}
	+O(\zeta^{-2}),
	\end{equation}
\end{enumerate}

\end{document}